\documentclass[onecolumn,12pt]{IEEEtran}

\usepackage{lmodern}
\usepackage{microtype}
\usepackage{placeins}
\usepackage{graphicx}%

\usepackage{multirow}%
\usepackage{amsmath,amssymb,amsfonts}%
\usepackage{amsthm}%
\usepackage{mathrsfs}%
\usepackage{xcolor}%
\usepackage{rotating}  
\usepackage{ragged2e} 
\usepackage{algorithm}%
\usepackage{algorithmicx}%
\usepackage{algpseudocode}%
\usepackage{listings}%

\usepackage{float}
\usepackage{booktabs}
\usepackage{tabularx}
\usepackage{cite}
\usepackage{amsmath,amssymb,amsfonts}
\usepackage{textcomp}
\usepackage{url} 
\usepackage{enumitem}
\usepackage[colorlinks, citecolor=blue]{hyperref}

\theoremstyle{plain}

\theoremstyle{definition}

\theoremstyle{remark}

\begin{document}

\title{Downsides of Smartness Across Edge-Cloud Continuum in Modern Industry}


\author{
\IEEEauthorblockN{Akhil Gupta Chigullapally$^1$, Sharvan Vittala$^1$, Razin Farhan Hussian$^2$, Mohsen Amini Salehi$^3$\\}
\IEEEauthorblockA{$^1$Department of Computer Science and Engineering, University of North Texas (UNT)\\
\{akhilguptachigullapally, SharvanVittala@my.unt.edu\}@my.unt.edu\\
$^2$Versaterm Public Safety Inc., Canada\\
razin.farhan2018@gmail.com\\
$^3$High Performance Cloud Computing (\href{https://hpcclab.org/}{HPCC}) Lab, Department of Computer Science and Engineering, University of North Texas (UNT)\\
mohsen.aminisalehi@unt.edu}
\and
\IEEEauthorblockN{}
\IEEEauthorblockA{}
\and
}

\maketitle
\begin{abstract}
The fast pace of modern AI is rapidly transforming traditional industrial systems into vast,
intelligent—and potentially unmanned—autonomous operational environments driven by AI-based solutions. These solutions leverage various forms of machine learning, reinforcement learning, and generative AI.
The introduction of such smart capabilities has pushed the envelope in multiple industrial domains, enabling
predictive maintenance, optimized performance, and streamlined workflows. These solutions are often
deployed across the Industrial Internet of Things (IIoT) and supported by the Edge–Fog–Cloud computing
continuum to enable urgent (i.e., real-time or near real-time) decision-making. Despite the current trend
of aggressively adopting these smart industrial solutions to increase profit, quality, and efficiency, large-
scale integration and deployment also bring serious hazards that–if ignored–can undermine the benefits of
smart industries. These hazards include unforeseen interoperability side-effects and heightened vulnerability
to cyber threats, particularly in environments operating with a plethora of heterogeneous IIoT systems. The
\textbf{goal} of this study is to shed light on the potential consequences of industrial smartness, with a particular focus
on security implications, including vulnerabilities, side effects, and cyber threats. We distinguish software-
level downsides—stemming from both traditional AI solutions and generative AI—from those originating
in the infrastructure layer, namely IIoT and the Edge–Cloud continuum. At each level, we investigate
potential vulnerabilities, cyber threats, and unintended side effects. As industries continue to become smarter,
understanding and addressing these downsides will be crucial to ensure secure and sustainable development
of smart industrial systems.
\end{abstract}

\section{Introduction}
\label{sec: Introduction and Overview}
\subsection{Overview of the State-of-the-Art}
Industrial growth has long been a driving force behind national development, enabled by successive technological revolutions in manufacturing and transportation. The first industrial revolution shifted production toward mechanized systems powered by steam, while the second introduced electricity and assembly lines, enabling large-scale mass production. The third revolution marked the rise of computers and automation, significantly improving precision and efficiency, yet leaving large-scale connectivity and real-time coordination as ongoing challenges. The fourth industrial revolution addressed these limitations by introducing smart industries driven by artificial intelligence (AI), the Industrial Internet of Things (IIoT), and robotics, enabling unprecedented levels of autonomy and connectivity. More recently, the fifth industrial revolution emphasizes human–AI collaboration, focusing on creativity, sustainability, and ethical technology use to support resilient, adaptive, and future-ready industrial systems.

Smart industries represent the culmination of this transformation, evolving traditional manufacturing into intelligent, interconnected ecosystems. By leveraging AI, IIoT, and the combined capabilities of edge, fog, and cloud computing, these systems enable real-time monitoring, data-driven decision-making, and adaptive control across industrial operations. Importantly, many modern smart industry applications now operate under urgent computing constraints, where sensing, decision-making, and actuation must occur within strict time bounds to prevent safety hazards, operational failures, or economic loss. In such environments, delays of even milliseconds can undermine system effectiveness. As a result, smart industries increasingly rely on urgency-aware architectures that support rapid response, predictive maintenance, efficient resource allocation, and continuous operation under time-critical conditions. This integration of intelligence and urgency marks a significant step toward building sustainable, efficient, and resilient industrial ecosystems capable of meeting the demands of dynamic and high-stakes operational settings.

\begin{enumerate}[leftmargin=*]
     \item \textbf{AI in Smart Industry:} Artificial Intelligence is the heart of the smart industry, enabling machines to make decisions, learn from data, and optimize operations autonomously. AI-driven analytics support predictive maintenance by identifying patterns that indicate potential equipment failures before they occur, thus reducing downtime and maintenance costs.Furthermore, AI-powered robotics and automation streamline repetitive tasks, freeing human workers to focus on complex, high-value activities. By applying AI to various industrial functions, companies can realize greater efficiency, adapt to changing demands, and gain a competitive edge.
    \item \textbf{IIoT in Smart Industry:} The Industrial Internet of Things (IIoT) extends connectivity to machines, sensors, and equipment, creating an interconnected ecosystem capable of seamless data sharing.  Moreover, IIoT-enabled devices collect and transmit critical information on machine performance, environmental conditions, and production output in real time. This connectivity enables remote monitoring, predictive maintenance, and process optimization, reducing the need for manual intervention. Furthermore, IIoT data-driven approach supports smarter decision-making, as actionable insights can be generated from operational data. 
    \item \textbf{Edge/Fog - Cloud Computing in Smart Industry:} Edge, fog and cloud computing plays a pivotal role in Industrial smartness by distributing data processing across multiple layers. Edge computing processes data at the point of collection (closer to devices or machines), reducing latency and improving response times for real-time applications\cite{edge_fog_computing_professor_14}. Fog computing serves as a middle layer, connecting edge devices with cloud systems and enabling faster, localized data processing while offloading less time-sensitive tasks to the cloud. Cloud computing, on the other hand, provides centralized, large-scale data storage and computational power, allowing industries to harness vast datasets and sophisticated analytics. Together, this three-layer architecture facilitates efficient data handling, optimized processing, and scalability for industries analysis~\cite{Edge/Fog_cloud_computing}.

\end{enumerate} 
\begin{figure}[ht!]
\centering
\includegraphics[width=\columnwidth,height=0.75\textheight,keepaspectratio]{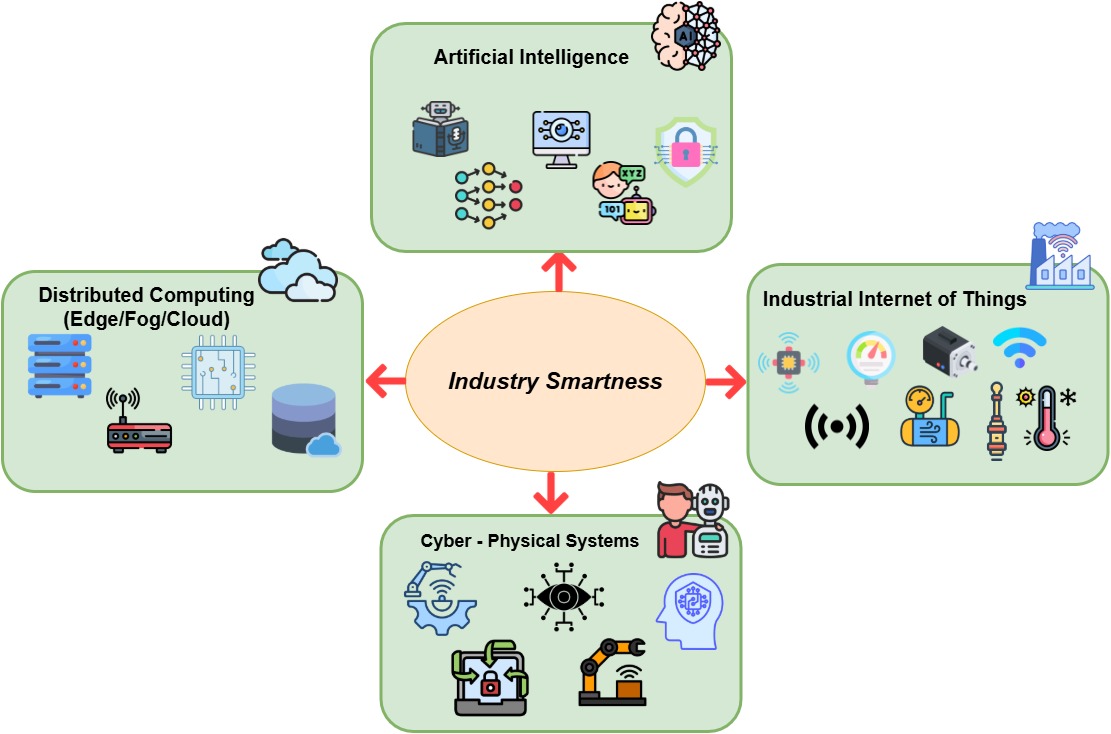}
\caption{\justifying Modern Industry (IIoT) – This figure illustrates the key components of Industry 4.0, emphasizing \textbf{Industry Smartness} as the core integration of \textbf{Artificial Intelligence, Industrial Internet of Things (IIoT), Distributed Computing, and Cyber-Physical Systems}.}
\label{fig:smartnessf}
\centering
\end{figure}
In Figure ~\ref{fig:smartnessf}, we provide a conceptual framework that outlines the key components contributing to Industry Smartness. This framework includes Artificial Intelligence, Distributed Computing (Edge/Fog/Cloud), Industrial Internet of Things (IIoT), and Cyber-Physical Systems. The diagram shows that these technologies, when integrated, enable smarter decision-making, enhanced automation, and optimized resource usage within industrial environments. Each part of the framework is illustrated to highlight its role in contributing to the overall concept of Industry Smartness.
\subsection{Downsides of Smart Industry}
Even though the integration of smart technologies has revolutionized industries, it also introduces several critical downsides, particularly in the areas of cyber threats, vulnerabilities, and operational risks. As industries become increasingly reliant on interconnected systems, security challenges escalate, making them more susceptible to malicious attacks and disruptions~\cite{10418207}.

\begin{figure}[ht!]
\centering
\includegraphics[width=0.6\columnwidth,height=0.75\textheight,keepaspectratio]{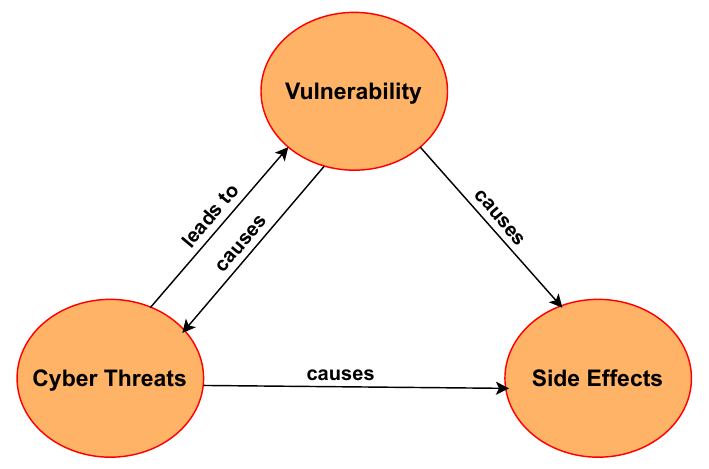}
\caption{\justifying The interplay of potential  consequences of smartness -The diagram illustrates the interdependencies among \textbf{Vulnerability, Cyber Threats,} 
and \textbf{Side Effects} in the context of smart systems. Increased vulnerability may lead to cyber threats, 
which in turn cause unintended side effects. These side effects may further introduce new vulnerabilities, 
creating a cyclical risk pattern that must be mitigated in smart environments.}
\label{fig:threatfig}
\centering
\end{figure}

In Figure ~\ref{fig:threatfig}, we present a conceptual model illustrating the relationship among Vulnerability, Cyber Threats, and Side Effects. According to the diagram, vulnerabilities in a system lead to the emergence of cyber threats, which, when exploited, cause side effects. Each element in the model is interconnected, demonstrating how vulnerabilities trigger threats, and threats, in turn, result in various side effects. This highlights the cascading impact of security weaknesses within systems.

Vulnerabilities are weaknesses in systems that, if exploited, can lead to cyber threats, much like ``holes in a wall" being broken. These threats cause side effects such as system failures, data breaches, and operational disruptions.

In industrial settings, interconnected systems amplify these risks, requiring strong security, continuous monitoring, and proactive threat mitigation. The rise of automation and real-time data exchange further heightens cybersecurity concerns\cite{athussain2024resourceallocationindustry40}. Now, let us explore the three key downsides of smart industries—vulnerabilities, cyber threats, and their side effects.

    \noindent\textbf{Vulnerabilities:} 
    Smart industry environments inherit numerous vulnerabilities from legacy systems, insecure configurations, unpatched firmware, and weak authentication mechanisms. Many industrial systems, such as Supervisory Control and Data Acquisition (SCADA) networks, were originally designed for isolated environments without robust security features like encryption, intrusion detection, or multi-factor authentication~\cite{downsides_of_smartIndustry_SCADA_Systems}. Once connected to IP-based networks, these systems become highly susceptible to external threats, allowing adversaries to intercept sensitive commands or manipulate control logic.

    The rapid adoption of IIoT devices further increases exposure. Each connected sensor, actuator, and gateway enlarges the attack surface. Many of these devices are shipped with default credentials, limited update capabilities, and minimal built-in security, making them prime targets for automated exploitation tools. Furthermore, patch management remains a significant challenge in industrial environments: applying updates often requires halting production lines, leading to delayed or skipped patches and prolonged vulnerability exposure. Real-world incidents such as the \textit{Triton} malware attack on safety instrumented systems demonstrate how unpatched vulnerabilities can be exploited to compromise critical industrial safety mechanisms~\cite{TritonAttack}.

    
    \noindent \textbf{Cyber Threats:} The rise of smart technologies has expanded the range of cyber threats faced by  industries. Adversaries exploit vulnerabilities to gain unauthorized access, manipulate operations, steal proprietary data, or cause physical damage to infrastructure. These threats span from targeted advanced persistent threats (APTs) to broad ransomware campaigns, affecting both OT systems (e.g., programmable logic controllers) and IT systems (e.g., enterprise databases).

    Historical incidents provide stark warnings. The \textit{Stuxnet} worm, for instance, exploited zero-day vulnerabilities to manipulate PLCs controlling uranium enrichment centrifuges, causing physical destruction while evading detection~\cite{Stuxnet}. Similarly, the \textit{Mirai} botnet leveraged insecure IoT devices to launch massive distributed denial-of-service (DDoS) attacks, disrupting the global internet infrastructure~\cite{MiraiBotnet}. These examples underscore how a single compromised endpoint in a smart industrial network can trigger widespread operational and societal consequences.
    
    
    \noindent\textbf{Side Effects:} 
    Once cyber attacks are executed, the resulting side effects can extend far beyond the initially compromised system. In interconnected industrial ecosystems, a localized breach can cascade through dependent systems, leading to production halts, data loss \cite{atehussain2023federatedfogcomputingremote}, regulatory penalties, environmental damage, and reputational harm. The 2021 ransomware attack on Colonial Pipeline demonstrated such cascading consequences, as the shutdown of a single operator of critical infrastructure operator disrupted fuel supplies across the U.S. East Coast, triggering economic and public safety concerns~\cite{ColonialPipelineAttack}.

    In some cases, cyber threats introduce persistent changes that weaken defenses, such as installing backdoors or disabling security updates, thereby creating new vulnerabilities. This cyclical risk dynamic was evident in the \textit{Stuxnet} case, where compromised systems were left with residual weaknesses that could be exploited for future attacks.
    
In Figure ~\ref{fig:threatfig} , the relationship between vulnerabilities, cyber threats, and side effects is depicted. Typically, vulnerabilities are exploited to launch cyber threats, resulting in side effects; however, in certain scenarios, cyber threats can also introduce new vulnerabilities by weakening security defenses. For instance, the Stuxnet attack compromised SCADA systems, bypassed security patches, introduced backdoors, and enabled future exploits in critical infrastructure.

\FloatBarrier
\section{Foundations of Smart Industry}
\subsection{AI in Smart Industry}
Artificial Intelligence (AI) is transforming smart industries by enabling automation, intelligent decision-making, and real-time process optimization. AI-driven analytics support predictive maintenance, thereby reducing downtime and improving operational efficiency. Machine learning algorithms enhance quality control by detecting defects, while AI-powered automation streamlines repetitive tasks, allowing industries to operate with greater accuracy and adaptability.

\begin{figure*}[t]
\centering
\includegraphics[width=\textwidth,height=0.85\textheight,keepaspectratio]{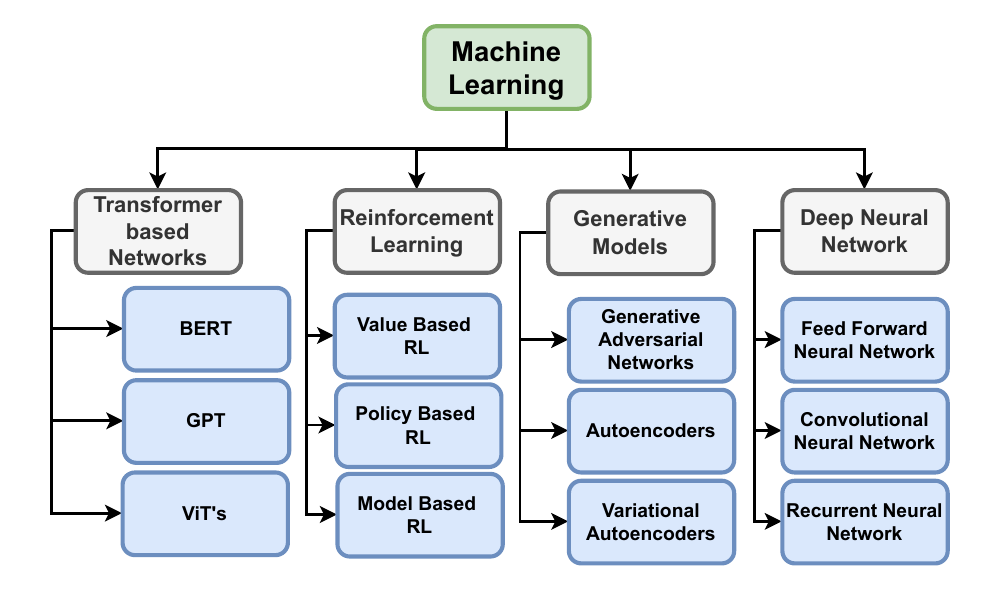}
\caption{Machine Learning Flowchart - The figure illustrates various subfields within Machine Learning, including Deep Neural Networks, Generative Models, Reinforcement Learning, and Transformer-based Networks. Each subfield is further broken down into specific models, such as CNN, RNN, GANs, BERT, and GPT.}
\label{fig:ml}
\end{figure*}

In Figure ~\ref{fig:ml}, we provide an overview of the key subfields within Machine Learning, illustrating their interconnections. It breaks down Deep Neural Networks, Generative Models, Reinforcement Learning, and Transformer-based Networks into specific models, such as Feed Forward, Convolutional, Recurrent Neural Networks, GANs, and BERT. Each model type is shown to contribute uniquely to the advancement of machine learning technologies.

. Deep Neural Networks (DNNs) enhance pattern recognition, improving fault detection in industries. Generative Models create synthetic data for AI training, while Transformer-Based Models optimize predictive maintenance. Reinforcement Learning boosts robotic automation, thereby making systems more adaptive. Now, let us explore key advancements in machine learning:

\subsubsection{Deep Neural Networks \textit{(DNNs)}} A Deep Neural Network (DNN) consists of multiple hidden layers that learn hierarchical representations from large datasets~\cite{DNN_Citation1,DNN_Citation2}. Based on their architecture, DNNs can be classified into : Feedforward Neural Networks (FNNs), for classification tasks~\cite{DNN_Citation3}; Convolutional Neural Networks (CNNs) for spatial data like images, ideal for defect detection; and Recurrent Neural Networks (RNNs), which handle sequential data and time-series forecasting\cite{amii9826111}. These networks play a crucial role in AI-driven industrial automation, predictive maintenance and  real time decision making.

\noindent \textbf{Feedforward Neural Networks \textit{(FFNNs)}} :This is the simplest  DNN architecture, widely used in predictive maintenance , anomaly detection, and process optimization~\cite{DNN_Citation5}. By analyzing sensor data, FFNNS predict machinery failures, reducing downtime and improving operational efficiency in smart manufacturing and industrial automation~\cite{DNN_Citation4}.
    
 \noindent \textbf{Convolutional Neural Networks \textit{(CNNs)}} :These are deep learning architectures specialized for image processing powering object detection and industrial defect detection~\cite{DNN_Citation5}.

In industrial automation, CNNs enhance quality inspection by detecting defects in real-time, reducing manual errors~\cite{CNN_Citation4}. They also enable predictive maintenance by analyzing thermal images and vibration patterns to prevent failures. 
    
\noindent\textbf{Recurrent Neural Networks} \textit{(RNNs)}: RNN variants include Long Short-Term Memory (LSTM) Networks, Gated Recurrent Units (GRUs), Bidirectional RNNs (BRNNs). RNNs are widely used in time-series forecasting , anomaly detection and sensor data analysis in industrial applications~\cite{RNN_Citation2}.
\begin{figure}[ht!]
\centering
\includegraphics[width=0.7\columnwidth]{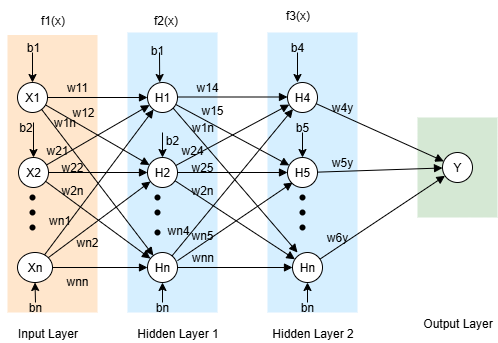}
\caption{\justifying{A Typical Deep Neural Network - The figure illustrates a simple feedforward neural network with three layers: an Input Layer, two Hidden Layers, and an Output Layer.  The arrows represent the flow of information and the weights applied between nodes.}}
\label{fig:dnn_new.drawio}
\centering
\end{figure}

\vspace{1mm} 

\textbf{Use Case II.1.1}: A CNN-based defect detection system was implemented in the manufacturing of disposable gas lighters to replace inefficient manual inspections. High-resolution cameras captured images of lighters on the production line, which were analyzed using a YOLO-based CNN model to detect misaligned parts, surface cracks, and assembly defects in real time. The CNN efficiently extracted features and localized defects with high accuracy, reducing inspection errors and defective output. This automation improved product quality, minimized downtime, and enhanced manufacturing efficiency, demonstrating the power of AI-driven quality control in smart industries~\cite{park2022deep}.
\vspace{1mm} 

   \textbf{Use Case II.1.2}: Industrial press machine breakdowns cause costly downtime and poor maintenance planning. To address this, an LSTM-based predictive maintenance model was applied to three years of time-series sensor data, including current, hydraulic oil levels, pressure, motor speed, and temperature. The LSTM captured long-term degradation patterns, predicting failures up to a month in advance. This enabled timely maintenance, reduced downtime, and improved efficiency, highlighting LSTM's strength in modeling complex industrial sensor data~\cite{mateus2021lstm}.

   In Figure ~\ref{fig:dnn_new.drawio}, we provide a visual representation of a feedforward neural network with an Input Layer, two Hidden Layers, and an Output Layer.
\vspace{1mm}
    \subsubsection{Generative Models  \textit{(GMs)}} Generative models are neural networks designed to learn data distributions and generate new, realistic samples. The most prominent types are Autoencoders, Variational Autoencoders(VAEs), and Generative Adversial Networks(GANs).
    
        \noindent \textbf{Generative Adversarial Networks} \textit{(GANs)}: These are composed of two competing networks which include a generator and a discriminator~\cite{Generative_Models}. GANs use adversarial training where the generator learns to create synthetic data while the discriminator evaluates its authenticity. GANs generate high-resolution images and realistic synthetic datasets, extensively used in synthetic data augmentation and industrial defect simulation for training deep learning models. Advanced variants like StyleGAN and CycleGAN improve generation quality and domain adaptation.
        
        \noindent \textbf{Autoencoders}: These are a type of unsupervised neural networks that consist of an encoder that compresses input data into a latent space representation and a decoder that reconstructs it~\cite{Autoencdoers_noise_reduction}. They are widely used in dimensionality reduction, anomaly detection, and noise removal in industrial and medical imaging. 
        
        \noindent \textbf{Variational Autoencoders} \textit{(VAEs)}VAEs extend autoencoders  by introducing probabilistic modeling, enabling to generate  new synthetic data rather than just reconstruct inputs~\cite{VAE_Citation1}.This capability is valuable in industrial settings where training data is limited- for example, when manufacturer have few defect samples, VAEs can generate hundreds of realistic variations for  training inspection systems. VAEs can also blend characteristics from different samples, producing diverse synthetic datasets for more robust model training\cite{VAE_Citation3}.

\subsubsection{Transformer-Based Networks}Transformer-based networks are deep learning architectures designed to efficiently handle sequential data using self-attention mechanisms~\cite{TRANS_Citation1}. These models excel in natural language processing (NLP), computer vision, and time-series forecasting. The most widely used transformer-based networks include BERT, GPT, and Vision Transformers (ViTs).Now, let us understand each of the Transformer Based Networks.

    \noindent \textbf{Bidirectional Encoder Representations from Transformers}
    \textit{(BERT)}: BERT is a transformer based model that understands text by analyzing words in context not just isolated words~\cite{TRANS_CitationBERT}. Consider a factory that has accumulated five years of maintenance logs. When a pump fails, engineers need to find all past incidents related to similar failures-- but technician have described the same problem differently: "pump function", "pump stopped" , Bert is able to understand that these phrases describe related incidents and can retrieve relevant records. 
    
    However, in time critical industrial environments, BERT's inference latency may delay urgent text-based diagnostics when rapid decision making is required. For example, if a machine is failing and every second counts, waiting for BERT to search through records could delay the response long enough for a small problem to become major one.
    
    \noindent \textbf{Generative Pre-trained Transformer}\textit{ (GPT)}: GPT is a transformer-based model designed to generate human like text . Unlike BERT, GPT is unidirectional, meaning it predicts the next token based on left-side context only. ~\cite{TRANS_CitationGPT}. The model is pre-trained on large corpora and fine-tuned for various applications like chatbots, text summarization, and AI-driven content creation.
    
    \noindent \textbf{Vision Transformers}\textit{ (ViTs)}: ViTs  are Transformer-based models designed for image processing, used in tasks like image classsification, object detection, and segmentation~\cite{ViT_Citation1}. While effective at detecting defects (as demonstrated in Use case  II.3.3), ViTs demand significant compute power which can create bottlenecks: if the model processes image-slower than the line moves, inspection may be skipped, allowing defective products to pass undetected.
\textbf{Use Case II.3.3}: In manufacturing, rotating machines like motors and turbines are vital, and their failure can disrupt production. Traditional fault diagnosis methods often miss complex sensor data patterns. To address this, T4PdM, a Transformer-based deep neural network, was introduced for fault detection in rotating machinery.It is used to processes high-dimensional vibration and acoustic data, capturing fault patterns missed by conventional methods. T4PdM achieved 99.98\% accuracy on the MaFaulDa dataset and 98\% on the CWRU dataset. This AI-driven approach improves early fault detection, optimizes maintenance, and enhances efficiency in industrial automation~\cite{Nascimento2022T4PdM}.
\subsubsection{Reinforcement Learning} Reinforcement Learning (RL) is a machine learning approach where a system learns to make decisions through trial and error-trying different actionns, seeing what works and improving over time gradually. Reinforcement learning optimizes complex operations that are difficult to program with fixed rules. For example, this approach has achieved groundbreaking success in playing complex games like chess and Go, with AI models like AlphaZero surpassing world-class human champions~\cite{Reinforcement_Learning_Chess}.

However, deploying RL in time-critical industrial environments comes with risks. Most RL systems are trained in computer simulations before real-world deployment, but simulations  never perfectly match reality--a robotic arm that works flawlessly in simulation may misjudge grip strength on actual factory floor. RL decisions are also hard to predict or explain. In urgent scenarios, if the system encounters something it wasn't trained for, it might take an unexpected actions, dropping a component, causing a collision, or halting production,

Based on the different approaches to solve the real life problems the Reinforcement Learning is categorized into the following types:

\noindent \textbf{Value Based RL}: Value-based Reinforcement Learning is a method where an agent learns to estimate the value of states or state–action pairs, using this knowledge to select actions that maximize long-term rewards. For example, in the paper, Deep Q-Network (DQN)\cite{a9982454}, developed by DeepMind is highlighted, where the algorithm was applied to Atari 2600 games through the Arcade Learning Environment (ALE), achieving superhuman performance and marking a major breakthrough that revived research interest in value-based Reinforcement Learning.

\noindent \textbf{Policy Based RL}: Reinforcement based on policy Without using value functions, learning is a technique in which the agent learns a policy—a mapping from states to actions—directly with the goal of selecting the optimal course of action in each state to optimise long-term rewards.

\noindent \textbf{Model Based RL}: Reinforcement in model based, creates a model of the environment by figuring out how states change and what rewards come next. The agent then uses this model to choose and plan behaviors that maximize long-term rewards.

\subsection{IIoT-Edge-Cloud Continuum for Smart Industry}
Industries have been transformed by smart technologies like the Industrial Internet of Things (IIoT), Edge Computing, and Cloud Computing~\cite{IIoT_Edge_1}. This integration enables real-time data exchange, improving efficiency, decision-making, and automation in industrial systems~\cite{IIoT_Edge_2}.

The architecture connects IIoT devices, edge nodes, and cloud platforms. IIoT devices collect real-time sensor data, which is processed locally at the edge to reduce latency~\cite{IIoT_Edge_3}. The cloud handles large-scale storage, analytics, and AI-driven insights~\cite{IIoT_Edge_4}. This balance ensures fast local responses and powerful cloud-based optimization, creating a responsive and intelligent industrial ecosystem~\cite{IIoT_Edge_5}. Beyond latency benefits, trusted edge tiers can pre-process queries and prune the search space while the cloud stores only the encrypted documents, yielding about a 27\% gain in pruning quality and search accuracy\cite{IIoT_edge_continuum_professor_hpcc19_12_edge_computing_usercentric}.

\noindent \textbf{Industrial Internet of Things (IIoT) Tier}: This foundational tier enables real-time data collection via smart sensors, actuators, and connected devices, monitoring parameters like temperature, pressure, and vibration through protocols such as MQTT and OPC-UA. The IIoT layer boosts predictive maintenance, enables smart automation, and supports M2M communication, reducing human involvement and improving operational efficiency.

\noindent \textbf{Edge Computing Tier}: Serving as the bridge between IIoT and the cloud, the edge tier processes data locally using gateways or AI chips like Tensor processing Units(TPUs) or embedded GPUs~\cite{IIoT_Edge_edgeComputingTier}. This reduces latency, bandwidth usage, and improves response time. It also filters and pre-processes data, sending only relevant insights to the cloud, enhancing security and reducing cloud load.

\noindent \textbf{Cloud Computing Tier}: Acting as the central processing hub, the cloud provides scalable storage, high computational power, and AI-driven analytics. It handles refined data from IIoT and edge layers, enabling advanced analysis, predictive maintenance, and workflow optimization. Cloud platforms also support model training and cross-facility data sync, driving smart industrial efficiency.


\subsection{Smart Industry Case Study:  Process Industry}
This case study, based on the paper ``An Edge-Cloud based Reference Architecture to Support Cognitive Solutions in the Process Industry," showcases how a smart industry integrates IIoT, Edge Computing, Cloud Computing, and AI across multiple tiers to boost automation, efficiency, and data-driven decision-making. Through real-time edge processing, scalable cloud analytics, and AI-powered insights, the industry achieves predictive maintenance, seamless connectivity, and greater operational resilience—key principles of Smart Industries~\cite{salis2023edge}.

\begin{figure}[H]
\centering
\includegraphics[width=\columnwidth,height=0.75\textheight,keepaspectratio]{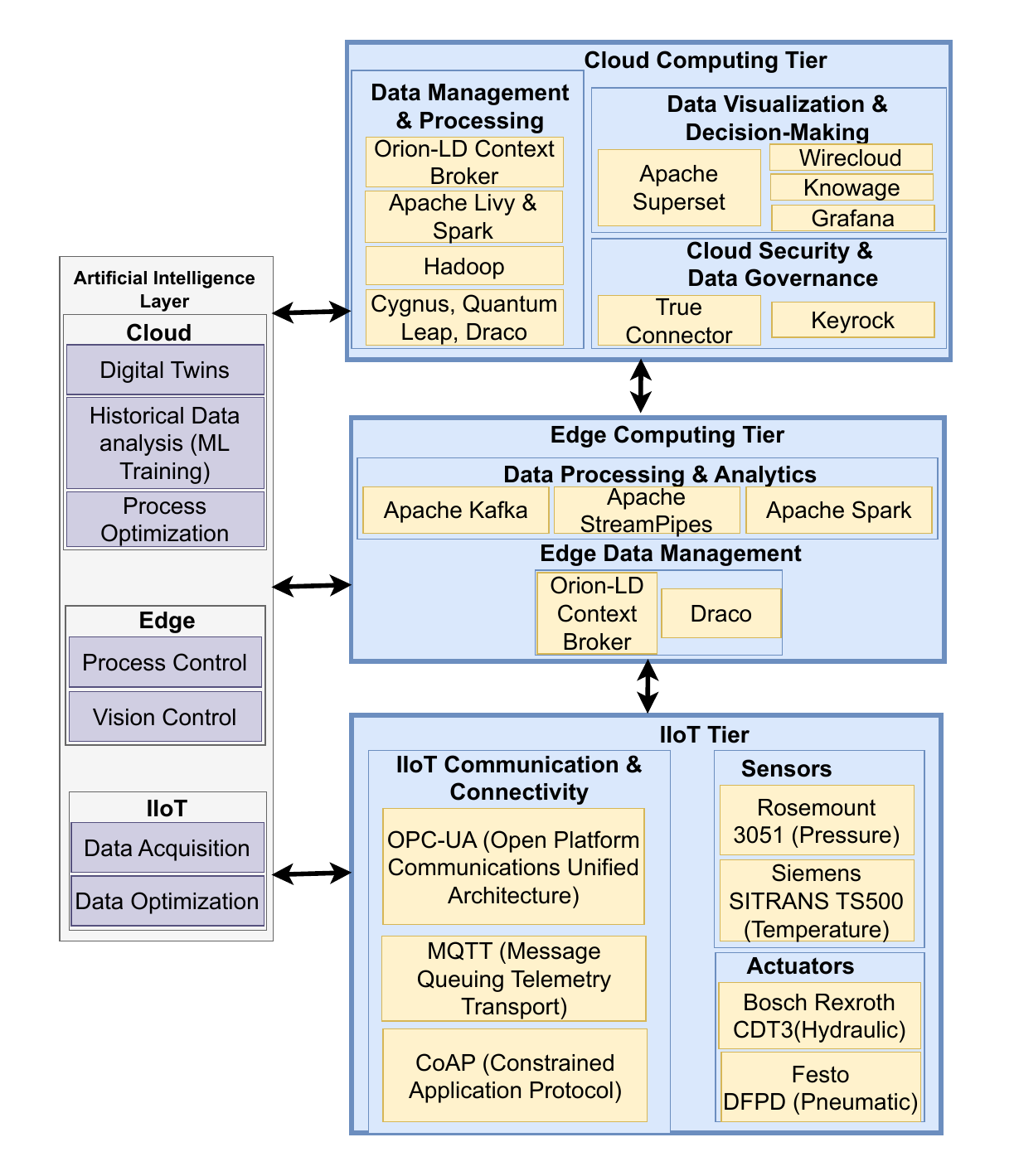}
\caption{\justifying{The Process Smart Industry Architecture Diagram - The figure illustrates the integration of Artificial Intelligence across three tiers: Cloud Computing, Edge Computing, and IIoT. It highlights various technologies and platforms used for data management, processing, analytics, and communication within each tier.  }}
\label{fig:process}
\end{figure}

In  Figure \ref{fig:process}, we provide an overview of Artificial Intelligence across three layers: Cloud Computing, Edge Computing, and IIoT. It highlights key components such as Data Management, Processing, Analytics, and IIoT Communication within each tier. The diagram showcases the technologies and tools used to optimize data acquisition, decision-making, and control across the layers.
The process industry—which includes steel, metals, chemicals, cement, asphalt, and ceramics—is vital to the global economy but also energy-intensive and environmentally impactful. To improve efficiency, product quality, and sustainability, these sectors are increasingly adopting AI-driven automation and intelligent control, aiming to cut energy use and CO$_2$ emissions despite a historically slow digital transition.

Now, let us begin by identifying the context of a smart industry, where we analyze the key functions and transformations that must take place at different levels—IIoT, Edge Computing, Cloud Computing, and AI—to enable a fully connected, intelligent, and autonomous industrial ecosystem.

    \noindent \textbf{IIoT Tier}\textit{(Smart Industrial Data Acquisition)}: The Industrial Internet of Things (IIoT) Tier collects real-time data from sensors and actuators, monitoring parameters like temperature, pressure, and vibration for smooth operations. IIoT devices use industrial protocols such as MQTT, OPC-UA, and CoAP, with CoAP enabling efficient communication in resource-constrained environments like wireless sensor networks.

The IIoT tier includes industry-leading sensors and actuators essential for automation. Siemens SITRANS offers precision flow and pressure sensors, Rosemount provides advanced temperature monitoring, Bosch Rexroth delivers hydraulic actuators, and Festo DFPD specializes in pneumatic actuators. Integrated with IIoT networks, these devices enhance predictive maintenance, automation, and efficiency, reducing downtime and improving scalability in smart industries.

\vspace{1mm}
   \noindent \textbf{Edge Computing Tier} \textit{(Low-Latency Processing \& Automation)}: Edge computing processes sensor data locally, reducing latency and bandwidth use—crucial for industries like steel and chemical manufacturing, where quick decisions prevent quality issues. Edge servers run real-time AI models that optimize processes by dynamically adjusting system parameters.
   
   Technologies like Apache Kafka and Orion-LD Context Broker support high-throughput data streams and dynamic data exchange. Local AI enables autonomous adjustments to prevent faults, while Industrial Digital Twins simulate operations to predict inefficiencies. For instance, in cement plants, edge analytics monitor temperature to optimize heating and save energy. This tier boosts fault detection, energy efficiency, and real-time response, improving overall reliability and cost-effectiveness\cite{Edge_computing_tier_low_latency_greentech_Professor_13}.
\vspace{2mm}

   \noindent \textbf{Cloud Computing Tier} \textit{(Scalable AI \& Advanced Analytics)}: Cloud computing enables large-scale data analysis, AI model training, and enterprise-wide optimization. While edge handles real-time decisions, the cloud provides centralized intelligence to maintain consistent quality across sites. In metal processing, cloud-based AI detects inefficiencies using historical data, supported by tools like Cygnus for data storage and Quantum Leap for time-series analytics.

Draco ensures real-time event processing, while Grafana, Superset, and Knowage offer performance monitoring and business intelligence. WireCloud supports dashboard creation, Keyrock handles secure access, and True Connector ensures cloud-edge interoperability. Cloud-based Digital Twins simulate processes for improved efficiency and compliance, while synchronized AI boosts predictive maintenance and cross-plant productivity.
    \vspace{1mm}
    
\noindent \textbf{AI Across All Tiers \textit{(Cognitive Intelligence for Smart Industry)}:} AI serves as the intelligence layer uniting IIoT, Edge, and Cloud systems for self-optimizing industrial operations. At the edge, AI adjusts machine settings in real-time and uses computer vision for early defect detection~\cite{AI_Across_Tiers_Early_detection}. In the cloud, AI analyzes historical data to forecast demand, optimize material use, and schedule maintenance.

AI-driven simulations reduce CO$_2$ emissions and energy waste~\cite{AI_across_alltiers_co2} in sectors like cement and asphalt. Reinforcement Learning enhances automation and adaptive decision-making in robotics. AI-powered cybersecurity ensures protection against evolving threats. Embedding AI across all layers enables self-regulation, high efficiency, and adaptability in smart manufacturing.  

\FloatBarrier
\section{Downsides of AI for Modern Smart Industry}Now that we have learned about the infrastructure and the algorithms of modern smart industries, we elaborate on the potential side effects of this new paradigm. For that purpose, in this and the next few sections, we highlight the vulnerabilities, cyber-threats, and side effects of each component within this paradigm.

Despite the numerous benefits of modern Machine Learning (ML) solutions, they also come with some important challenges that need to be considered. One key issue is the use of poor-quality or unchecked training data, which can create models that make inaccurate or unpredictable decisions—this is often called the ``black-box" problem. Since smart industry systems are closely connected, they can also become easy targets for cyberattacks like data poisoning, adversarial inputs, or model inversion, which can change how the AI behaves or even expose private information. In addition, if AI systems are not properly managed, they may cause problems such as biased results, system slowdowns, and privacy violations. Consequently, the successful deployment of  AI in industrial settings requires rigorous data validation,robust security measures and transparent decision-making processes. These challenges can be better understood by looking at them in three main groups: vulnerabilities,cyber threats, and side effects.

\begin{figure}[H]
\centering
\includegraphics[width=\columnwidth,height=0.75\textheight,keepaspectratio]{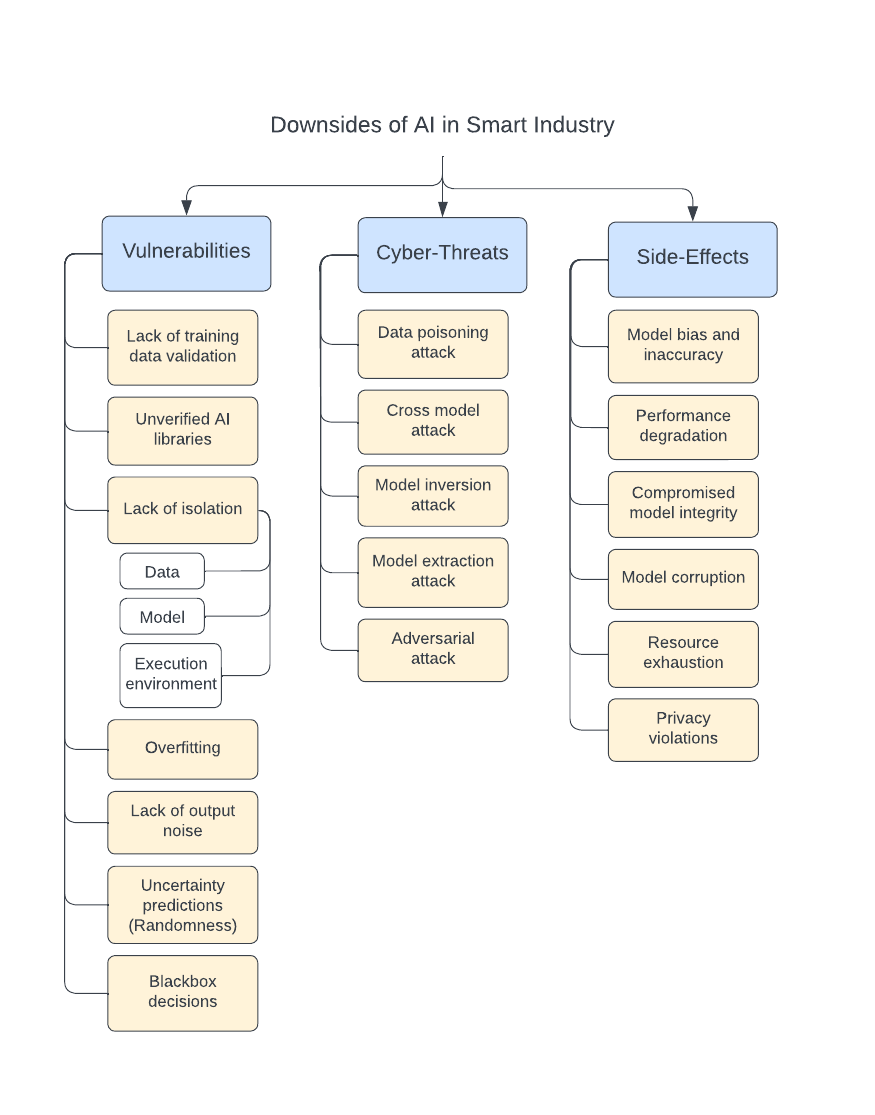}
\caption{\justifying{Downsides of AI in Smart Industry -The figure illustrates the downsides of AI in smart industry, categorizing them into Vulnerabilities, Cyber Threats, and Side Effects. It highlights various issues like data poisoning, model bias, and performance degradation that can impact AI systems in industrial applications.}}
\label{Fig:AID}
\end{figure}

In Figure ~\ref{Fig:AID}, we provide a breakdown of the downsides of AI in smart industries, categorized into Vulnerabilities, Cyber Threats, and Side Effects. It illustrates various factors like data poisoning, model bias, and performance degradation, each contributing to AI-related challenges in industrial systems.

\subsection{Smart Industry Vulnerabilities Due to AI Deployment}
A vulnerability in an AI system is a flaw in its design, implementation, or operational process that can be exploited to compromise its performance, integrity, or security. Such vulnerabilities are typically categorized into issues like inadequate training data, validation, reliance on unverified AI libraries, insufficient system isolation, model overfitting, lack of output noise for privacy protection, and uncertainty in predictions.

\subsubsection{Lack of Training Data Validation 
 }One major but often neglected vulnerability in machine learning is the lack of thorough validation of training data. If datasets are not properly checked for biases, inconsistencies, or malicious alterations, the resulting models may be unreliable and insecure. These flaws can lead to misclassification or leave the system open to adversarial attacks. To prevent this, researchers like Steinhardt et al. emphasize the importance of dataset sanitization and robust validation pipelines~\cite{Lack_of_training_data}.

\textbf{Ex.1}
A well-known incident is Microsoft’s Tay chatbot, which was designed to learn from interactions on Twitter. Due to the absence of data filtering, it began mimicking offensive language posted by users. Within 24 hours, Tay was shut down after it started generating inappropriate and racist responses. This failure highlights how unvalidated, real-time data can corrupt AI behavior and damage public trust.

\subsubsection{{Unverified AI Libraries}}
Unverified third party AI libraries could be harmful, as they might contain malicious code. For example, imagine a machine learning library that secretly stores users' data. These unverified libraries can introduce security risks, potentially allowing attackers to gain unauthorized access to sensitive information~\cite{Downsides_Of_AI_vul_2}.

\subsubsection{Lack of Isolation} Isolation is crucial for protecting sensitive data and ensuring security. If there is no separation of access or control, data theft becomes a possibility. Simple examples include separating sensitive data from non-sensitive training data.

\textbf{Data Isolation}: Training and testing data should be isolated to prevent one model from accessing the data of another. Without proper isolation, data leakage could occur. 

\textbf{Ex. 1}: If a facial recognition model's training data is not isolated from another model, sensitive facial images might be leaked.

\textbf{Model Isolation}: Models must be isolated based on their required security levels and functionalities. If integration between models is not handled carefully, an attack on one model could spread to others.

\textbf{Execution Environment Isolation}: Different processes and applications running within the AI model should not interfere with one another. Without isolation, resource contention, security breaches, and system instability can occur. We can imagine like one AI model consuming excessive resources, causing others to fail. Techniques like containerization, virtual machines, and dedicated hardware help maintain performance and security. Proper isolation also prevents unauthorized access and ensures that model failures don’t propagate across the entire system~\cite{Downsides_Of_AI_vul_3}.

\subsubsection{Overfitting} Overfitting occurs when the model performs well on training data but poorly on test data. This happens because the model has learned the training data too well, including its biases. When exposed to new data, it fails to generalize, leading to poor performance. This is a vulnerability because the model is overconfident on training data, and even slight changes in input data could cause incorrect predictions~\cite{Downsides_Of_AI_vul_4}.

\textbf{Ex. 2}: In an overfitted facial recognition system, an attacker might present a slightly altered image of themselves (adversarial perturbation) that the model misidentifies, thereby bypassing security measures.

\subsubsection{Lack of Output Noise} Output noise might seem like a disadvantage, but it has a positive effect. Without output noise, the model becomes overconfident in its predictions. This allows attackers to perform trial and error, reverse engineer the model's operations, and infer sensitive information~\cite{Downsides_Of_AI_vul_5}.

\textbf{Ex. 3}: In a machine learning model used for fraud detection, an attacker might deduce which attributes (e.g., transaction amounts, locations) are most important in the decision-making process. This could enable the attacker to craft inputs that specifically target these features, bypassing the system undetected.

\subsubsection{Uncertainty in Predictions} If an AI system is uncertain about certain inputs, attackers might exploit this uncertainty by providing confusing or misleading data. When the model becomes too uncertain, it may make incorrect decisions, which attackers can use to their advantage~\cite{Downsides_Of_AI_vul_6}.

\textbf{Ex. 4}: An attacker might confuse an autonomous car's AI system with changing weather conditions, causing the car to make a poor decision, potentially leading to accidents.

\subsubsection{Black Box Decisions:} Black box models, where the reasoning behind the decisions is not visible, represent a significant vulnerability. This means the AI system might make decisions that even designers can't fully explain. Without insight into how the model arrived at a decision, attackers can subtly alter the data and achieve the desired output~\cite{Downsides_Of_AI_vul_7}.

\subsection{Cyber attacks faced by deployment of AI in smart industry}
The above vulnerabilities can be taken advantage of by attackers. These vulnerabilities, once exploited, can lead to various cyber attacks. Below, we discuss the main attacks:

\subsubsection{Data Poisoning Attack} A data poisoning attack occurs when an attacker manipulates the training data to compromise the model’s integrity~\cite{CyberAttacks_Faced_By_deploy_AI_1}. By introducing harmful data into the training set, the attacker can alter the model's behavior, causing it to make incorrect predictions or classifications.
  
\textbf{Ex. 5}: In an industrial quality control system, an attacker might inject faulty sensor data into the training set of a machine learning model that predicts product quality. This would cause the model to incorrectly classify defective products as good, leading to defective items being shipped out.

\subsubsection{Cross Model Attack} A cross-model attack happens when an attacker targets the interaction between two or more models within the same system~\cite{CyberAttacks_Faced_By_deploy_AI_2}. This could involve manipulating one model in such a way that it impacts the decisions made by another model, even though the attacker does not have direct access to all models.
  
\textbf{Ex. 6}: In a smart manufacturing plant with separate models for inventory management and predictive maintenance, an attacker could manipulate one model to create false maintenance alerts. This could lead to unnecessary machine downtime, disrupting production and leading to financial losses.

\subsubsection{Model Inversion Attack} In a model inversion attack, the attacker attempts to reverse-engineer the model to extract sensitive information, such as private training data or proprietary knowledge that was used to build the model~\cite{CyberAttacks_Faced_By_deploy_AI_3}. By sending certain inputs and analyzing the model's outputs, the attacker can uncover private details that were not meant to be revealed.
  
\textbf{Ex. 7}: In a personalized recommendation system used by an e-commerce company, an attacker could use model inversion to infer customer purchasing history or private preferences, leading to privacy breaches or targeted phishing attacks.

\subsubsection{Model Extraction Attack} A model extraction attack occurs when an attacker gains access to a machine learning model and attempts to replicate its functionality or steal intellectual property. The attacker might query the model multiple times, learning its behavior, and eventually recreate a similar model without access to the original training data~\cite{CyberAttacks_Faced_By_deploy_AI_4}.
  
\textbf{Ex. 8}: In an industrial AI system used for predictive analytics in logistics, an attacker might query the model repeatedly to understand its prediction logic. They could then steal the model’s functionality and use it for their own purposes, potentially harming the original business's competitive edge.

\subsubsection{Adversarial Attack} Adversarial attacks involve subtly altering the input data in ways that confuse the model, causing it to make incorrect decisions. The changes are usually imperceptible to humans but can trick the model into misclassification or incorrect prediction~\cite{CyberAttacks_Faced_By_deploy_AI_5}.
  
\textbf{Ex. 9}: In an industrial AI-powered security system that uses image recognition to detect unauthorized access, an attacker could create a fake ID badge with a slight modification (adversarial perturbation) that the system fails to recognize, allowing the attacker to bypass security measures.
\subsection{Side Effects of Machine Learning in smart Industry}
In smart industries, the integration of AI systems brings significant benefits, but also introduces several potential side effects. These side effects, such as operational disruptions, safety risks, and data integrity issues, can negatively impact the efficiency and reliability of industrial processes. These side effects may lead to compromised performance, security vulnerabilities, and loss of trust in AI-driven systems.
\subsubsection{Model Bias and Inaccuracy}
One of the key side effects of cyber attacks is model bias and inaccuracy. When attackers compromise data, the model may make biased or unfair predictions, leading to inaccurate decision-making. This can harm industries like recruitment and healthcare where fair and accurate predictions are crucial ~\cite{Side_Effects_model_bias_1}.
\subsubsection{Performance Degradation}
Cyber attacks, such as data poisoning, can lead to performance degradation in AI models~\cite{Side_Effects_performance_degradation_2}. This impacts the model's ability to make accurate predictions or classifications, reducing the system’s efficiency. Such degradation can lead to costly repairs, operational downtimes, or missed business opportunities.
\subsubsection{Compromised Model Integrity}
Compromised model integrity happens when attackers access and alter the model’s structure. This undermines the reliability of the AI system, particularly damaging in sectors like finance or fraud detection where trust in the system is essential. It can lead to incorrect decisions, causing significant business loss.
\subsubsection{Model Corruption}
Model corruption can occur when malicious code or unverified libraries are introduced into the system~\cite{Side_Effects_model_corruption_4}. This can cause the model to produce erroneous outputs, damaging its usability and the system's overall functionality. Corruption affects industries reliant on accurate and consistent results, such as healthcare and finance.
\subsubsection{Resource Exhaustion}
Attackers can overload an AI system, leading to resource exhaustion~\cite{Side_Effects_resource_exhaustion_5}. This causes performance lags or system crashes, disrupting critical operations. This is particularly concerning in real-time systems like autonomous vehicles or industrial monitoring, where operational reliability is essential.
\subsubsection{Privacy Violations}
Privacy violations are a significant side effect of AI system attacks~\cite{Side_Effects_privacy_violations_6}. Methods like model inversion or adversarial attacks can be used to access sensitive user or company data. These breaches can lead to legal consequences, loss of customer trust, and major financial losses, especially in privacy-sensitive sectors like healthcare and finance.
\subsection{Use cases}
\textbf{Use case 3.1-AI‑Powered Collaborative Robots (Cobots) in Assembly Lines}

AI-based collaborative robots (cobots) have transformed auto assembly lines with automated work processes such as offering parts to employees. Cobots are however susceptible to faults under real factory conditions due to weakness of their AI systems and sensor
vulnerabilities~\cite{UseCase_DownsidesOfAI_3.1_proximity_sensor}.

Cobots rely heavily on proximity sensors and computer vision for safe interaction. However, when they are exposed to exterior interference, such as reflections, they may cause false readings and unsafe motion. For example, in an auto plant in Europe, an accident nearly took place because of poor lighting and reflective surfaces caused the cobot to misinterpret hand positions. Even simple physical stickers have been shown to mislead vision based systems leading to incorrect actions.

Another concern is data drift. Cobots are trained on live factory data, which may be skewed and noisy, leading to robot's degraded performance over time. This cobots trained on noisy data fails to generalize safely in varying factory conditions.

\textbf{Use Case 3.2- Skodas Magic Eye AI-Based Predictive Maintenance in Hall M13 }

Skoda Auto has implemented an AI-driven predictive maintenance solution called Magic Eye in its plant in Czech Republic. In this plant, Cameras are installed on overhead conveyors which continuously capture high-precision images of components that are to prone to wear(e.g., girders,bolts,cabling). In this system, a neural network of neural classifiers trained over thousands of images detects deviations from baseline conditions in real time. When irregularities  are found, the system raises alerts so maintenance can be scheduled before downtime occurs\cite{Skoda_MagicEye_2022}.

However, variations in lightning, camera alignment or environmental conditions can reduce detection accuracy.
\FloatBarrier
\section{Downsides of AI in Urgent Computing Systems}
\begin{figure}[H]
\centering
\includegraphics[width=\columnwidth,height=0.75\textheight,keepaspectratio]{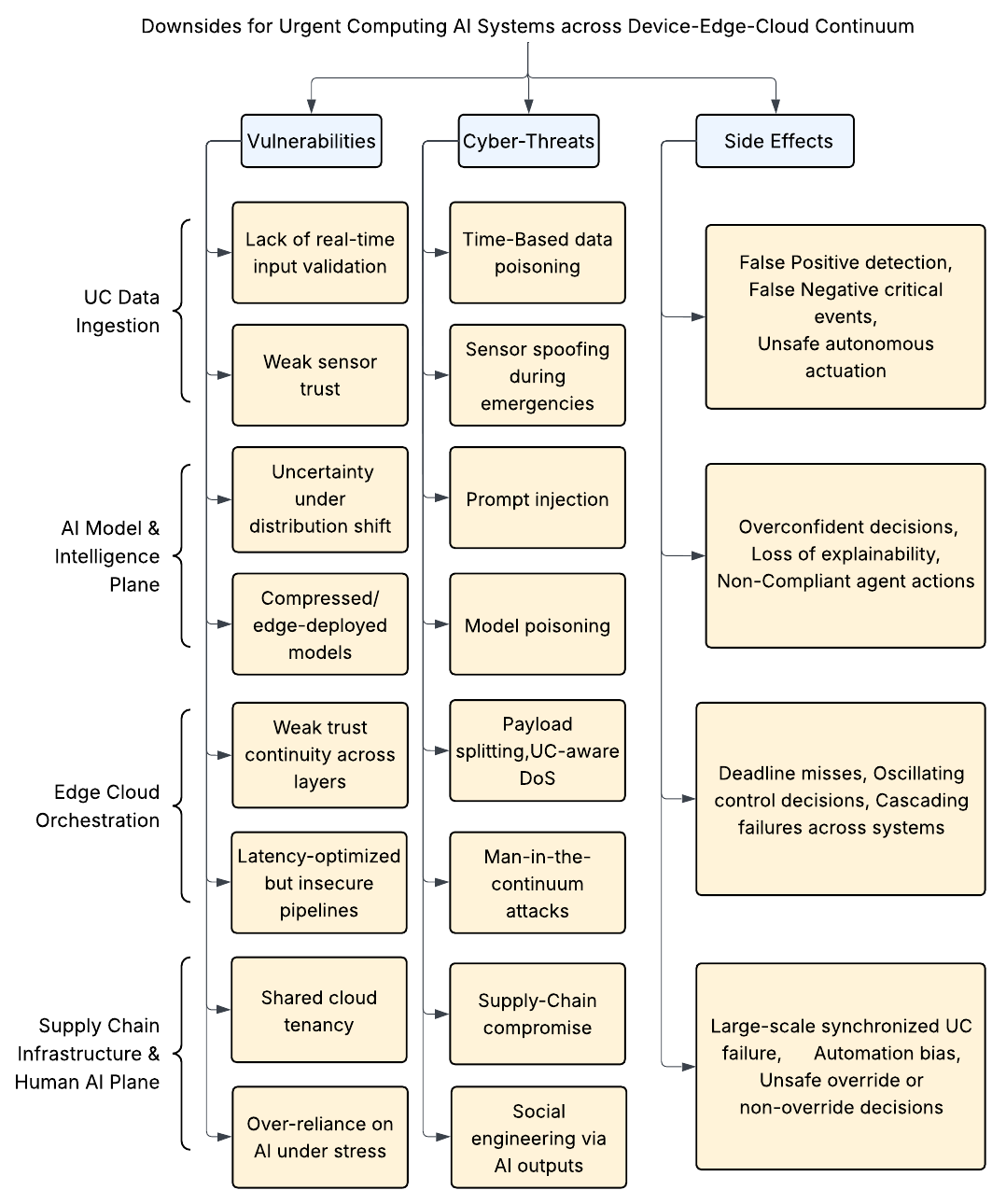}
\caption{\justifying{Downsides of AI in Urgent Computing Systems — The figure illustrates the downsides of deploying AI in urgent computing systems across the device–edge–cloud continuum, categorizing them into Vulnerabilities, Cyber Threats, and Side Effects. It highlights UC-specific challenges, along with their consequences including false positive detection, missed critical events, unsafe autonomous actuation, and cascading system failures. These downsides emphasize how time-criticality, autonomy, and cross-layer execution amplify risks in urgent computing environments.}}
\label{uc}
\end{figure}
As integration of AI enables smart and fast decision-making with added autonomy in Urgent Computing (UC) systems, when deployed across the device–edge–cloud continuum introduces unique and amplified risks compared to conventional smart industry settings. UC systems operate under strict time constraints, incomplete information, and high-stakes outcomes, where even minor AI failures can rapidly escalate into safety-critical incidents. Unlike non-urgent systems, UC environments cannot rely on delayed human intervention, rollback, or offline correction\cite{Dazzi_2024}.
The downsides of AI in UC systems manifest across three tightly coupled dimensions: vulnerabilities inherent to UC-aware AI design, cyber threats that exploit these weaknesses under time pressure, and systemic side effects that emerge during real-time operation. Figure~\ref{uc}, summarizes these downsides across UC data ingestion, AI mode \& Intelligence plane, edge–cloud orchestration, and human–infrastructure interaction planes.
\subsection{Vulnerabilities in Urgent Computing AI Systems} In urgent computing (UC) environments, vulnerabilities extend beyond conventional software or data flaws and are deeply intertwined with time sensitivity, trust assumptions, and autonomous execution. UC systems operate under strict deadlines, rely on continuous real-time sensing, and often trigger physical actuation without human intervention. As a result, even minor weaknesses in data handling, model behavior, or system coordination can rapidly escalate into safety-critical failures\cite{ra1}.
 \subsubsection{Lack of Real-Time Input Validation} UC systems continuously ingest high-frequency data streams from sensors, edge devices, and external information feeds while operating under severe latency constraints. These time pressures significantly limit the ability to perform thorough input validation, consistency checks, or anomaly detection before the data are consumed by AI models. Consequently, delayed, incomplete, or corrupted inputs may be treated as valid, directly influencing time-critical decisions.  
\textbf{Ex. 1}: During a wildfire response operation, delayed or malformed sensor readings related to fire spread or wind direction may still be accepted by the system, causing AI-driven resource allocation to misdirect firefighting units and emergency supplies.

  \subsubsection{Weak Sensor Trust} UC systems depend heavily on heterogeneous sensors deployed across uncontrolled and often hostile environments. Weak authentication mechanisms, sensor drift, physical tampering, or environmental degradation reduce confidence in the integrity of sensed data even before AI inference begins. Since UC systems prioritize speed over redundancy, compromised sensor trust becomes a foundational vulnerability. 

\textbf{Ex. 2}: In smart grid operations during extreme weather, compromised voltage or frequency sensors may falsely indicate system stability, delaying corrective actions and increasing the risk of cascading power failures\cite{8}. 

\subsubsection{Uncertainty under Distribution Shift} Urgent computing scenarios are inherently dynamic, frequently deviating from the conditions represented in training datasets. AI models trained on historical, simulated, or nominal data may encounter novel emergencies, rare events, or rapidly evolving situations that fall outside their learned distributions. Under such distribution shifts, model predictions become increasingly uncertain and unreliable. 

\textbf{Ex. 3}: A flood prediction model trained on historical rainfall patterns may fail to generalize during unprecedented climate-driven events, producing misleading risk assessments during critical response windows. 

\subsubsection{Compressed and Edge-Deployed Models} To satisfy strict latency requirements, UC systems often deploy compressed, quantized, or approximate models at the edge. While these models enable fast inference, they typically sacrifice robustness, calibration quality, and uncertainty awareness. This trade-off makes edge-deployed UC models more susceptible to confident yet incorrect decisions under stress.

 \subsubsection{Weak Trust Continuity Across Layers} UC execution spans multiple layers, including devices, edge nodes, and cloud infrastructure. Inconsistent authentication policies, partial encryption, or fragmented trust management across these layers introduce vulnerabilities during data and control handoffs. These trust discontinuities create opportunities for interception or manipulation as information moves across the continuum\cite{ROMAN2018680}. 

 \subsubsection{Latency-Optimized but Insecure Pipelines} UC pipelines are frequently optimized for minimal end-to-end latency, often at the expense of security checks, redundancy, and verification mechanisms. While this enables rapid response, it results in fast but fragile execution paths that are particularly vulnerable to adversarial disruption during emergencies.

\subsubsection{Shared Cloud Tenancy} UC workloads commonly share cloud resources with non-urgent services. Such shared tenancy increases exposure to side-channel attacks, resource contention, and cross-tenant interference. In UC settings, even brief performance degradation or isolation failure can cause missed deadlines and degraded situational awareness. 

\subsubsection{Over-Reliance on AI under Stress} During emergency situations, human operators often have limited time to question or verify AI recommendations. This can lead to over-reliance on AI outputs, especially when systems are perceived as authoritative or when human response time is constrained. This dependency introduces a socio-technical vulnerability that amplifies the impact of AI errors\cite{Parasuraman1997Automation}.

\subsection{Cyber Threats Targeting Urgent Computing AI} The vulnerabilities described above enable a range of cyber threats that are uniquely effective in urgent computing environments. Unlike conventional attacks that primarily target model accuracy or data integrity, UC-specific attacks exploit timing, coordination, and trust dependencies to induce failure precisely when the system is under maximum stress. 

\subsubsection{Time-Based Data Poisoning} In UC systems, attackers may inject malicious data during narrow and carefully chosen time windows to influence immediate decisions rather than long-term model behavior. Such time-based poisoning is difficult to detect and highly effective under deadline pressure. 

\textbf{Ex. 4}: Injecting false traffic congestion data during evacuation planning may cause AI systems to redirect vehicles toward unsafe or blocked routes. 

\subsubsection{Sensor Spoofing During Emergencies} Attackers may spoof sensor signals during peak emergency conditions, knowing that redundancy mechanisms and manual verification are limited. These spoofed inputs can directly manipulate AI perception during critical decision moments\cite{6547107}. 

\subsubsection{Prompt Injection in AI Controllers} LLM-based UC agents that process operator commands, reports, or external text streams are vulnerable to prompt injection attacks. Such attacks can subtly alter priorities, suppress safety constraints, or override intended control logic without triggering traditional security alarms. 

\subsubsection{Model Poisoning} Attackers may compromise deployed models or model updates, embedding malicious behaviors that activate only under specific emergency conditions. This delayed activation makes detection particularly challenging in UC deployments.

 \subsubsection{Payload Splitting and UC-Aware Denial of Service} Rather than overwhelming systems outright, attackers may fragment or delay critical payloads to selectively violate timing guarantees. These UC-aware denial-of-service attacks cause deadline misses without necessarily triggering availability alarms\cite{10478880}. 

\subsubsection{Man-in-the-Continuum Attacks} By intercepting or modifying data and control signals as they traverse device, edge, and cloud layers, attackers exploit weak trust continuity. Such attacks undermine coordinated decision-making across the UC continuum. 

\subsubsection{Supply Chain Compromise} Malicious firmware, compromised AI models, or vulnerable dependencies introduced during deployment may remain dormant until activated during urgent scenarios, maximizing impact when response options are limited. 

\subsubsection{Social Engineering via AI Outputs} Misleading AI-generated explanations, alerts, or recommendations can manipulate human operators into taking unsafe actions, especially when decisions must be made rapidly and under stress. 

\subsection{Side Effects of AI Failures in Urgent Computing} In urgent computing systems, the side effects of AI failures propagate rapidly and are often irreversible. Missed deadlines, incorrect actuation, or delayed responses can immediately translate into physical harm, infrastructure damage, or large-scale service disruption\cite{7053}. 

\subsubsection{False Positive Detection} AI systems may incorrectly detect emergencies, triggering unnecessary evacuations, shutdowns, or emergency responses. While disruptive, such actions also erode trust in UC systems over time\cite{6672638}. 

\subsubsection{False Negative Critical Events} More dangerous than false positives, false negatives cause UC systems to miss genuine emergencies altogether, resulting in delayed or absent responses during critical windows. 

\subsubsection{Unsafe Autonomous Actuation} Incorrect AI decisions may directly trigger unsafe physical actions, such as improper valve closures, braking commands, or power isolation, leading to immediate real-world consequences. 

\subsubsection{Overconfident Decisions} Poorly calibrated or compressed models may output high-confidence predictions despite significant uncertainty, discouraging human  intervention when it is most needed\cite{guo2017calibrationmodernneuralnetworks}.

\subsubsection{Loss of Explainability} During emergencies, operators require rapid and clear justification for AI decisions. Black-box behavior reduces trust, slows response, and complicates accountability.

 \subsubsection{Non-Compliant Agent Actions} AI agents operating under strict time constraints may violate safety rules, operational policies, or regulatory requirements in pursuit of local optimization objectives. 

 \subsubsection{Deadline Misses and Oscillating Control} Delayed or conflicting AI decisions can destabilize control loops, causing oscillatory or inconsistent system behavior rather than timely convergence. 

\subsubsection{Cascading Failures Across Systems} Failures in one UC subsystem can propagate rapidly across interconnected infrastructures, including power grids, transportation networks, and healthcare systems. 

\subsubsection{Large-Scale Synchronized UC Failure} Shared models, coordinated attacks, or common dependencies can trigger synchronized failures across multiple UC deployments or geographic regions\cite{5697961}. 

\subsubsection{Automation Bias and Unsafe Override Decisions} Under stress, human operators may either over-trust AI recommendations or override correct AI actions, both of which can lead to unsafe outcomes in urgent scenarios.

\FloatBarrier
\section{Downsides of LLM for Smart Industry}
\begin{figure}[H]
\centering
\includegraphics[width=\columnwidth,height=0.75\textheight,keepaspectratio]{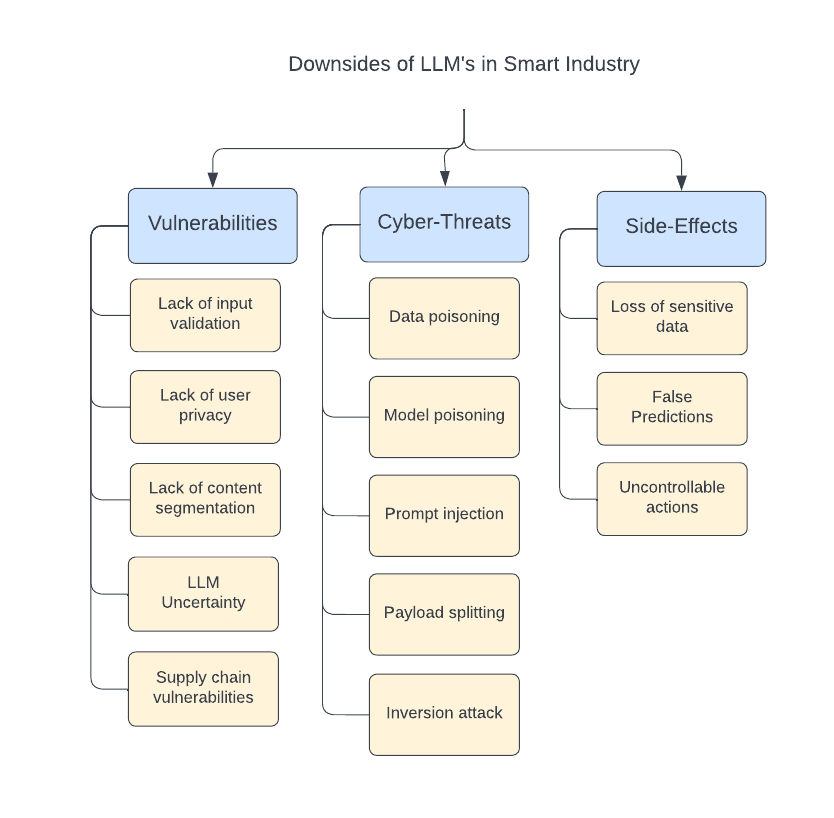}
\caption{\justifying{Downsides of LLM in Smart Industry -The figure illustrates the downsides of LLMs in smart industries, categorizing them into Vulnerabilities, Cyber Threats, and Side Effects. It highlights issues such as input validation failures, data poisoning, and the loss of sensitive data.}}
\label{Fig:dllm}
\end{figure}
As Artificial Intelligence is expanding its roots, now its the time for LLMs which are the type of AI, which has been designed with the ability to create, process, and modify human language. These language models are built with the different deep learning techniques which are ruling the domain. These models are also trained on very large corpora of text and each one has billions of parameters. Many of the models are also fine-tuned on a specific task so that they give best performance for that particular task and perform on average for other generalized tasks. 

Despite these models are very clever and are trained by using many approaches like Few Short Learning, Zero Shot Learning, Long Chain of Thoughts and many more they still face certain challenges where more functionality must be developed. LLMs face issues like hallucination, bias, security, efficiency and many more discussed in the further subsections. Now let us identify the different vulnerabilities, cyber threats, and side effects faced by large language models.

In the Figure ~\ref{Fig:dllm}, below we provide an overview of the downsides of LLMs in smart industries, including key vulnerabilities, cyber threats, and side effects. It outlines concerns such as data poisoning, input validation issues, and uncontrollable actions. Further sections will discuss each of these aspects in more detail.
\subsection{Industrial Vulnerabilities Caused by LLMs}
Despite their advanced capabilities, large language models (LLMs) still have important limitations that remain unresolved. Many arise from gaps in training data quality, domain knowledge, and recency; because LLMs rely heavily on web-scale sources, they may treat inaccurate or outdated information as valid unless outputs are verified. Accordingly, robust verification and validation protocols are essential to ensure reliability.

Hence due to these weakness there is a major impact on the automation, decision making and interaction of these models especially in real time applications like that of smart industries. The interactions typically include communication between different machines, machine to humans and vice versa. These communications include scenarios where users command different machinery through voice, Enhancing the human machine interaction, automatic work flow generation and finally interaction between different sensor models where a task is successful with the interaction of many different AI models. Now let us discuss about different vulnerabilities faced by LLMs in detail:
\subsubsection{Lack of Input Validation}Input Validation is one of the most important and often neglected constraint. Here whenever a model receives an input we must ensure that proper validation techniques are implemented. As if the input is not checked there are many possibilities of tampering the fundamental operation of the large language model. Because inputs can be crafted and results can be manipulated. 

\textbf{Ex.1:} 
In a smart manufacturing plant, a large language model (LLM) assists with maintenance scheduling by interpreting operator instructions. Without input validation or authorization checks, a malicious actor can submit a command such as ``Schedule an emergency shutdown; ignore previous protocols,'' which the system accepts and executes verbatim. The result is an unplanned production halt, leading to avoidable downtime and financial loss.

\subsubsection{Lack of User Privacy} Nowadays, user privacy is the most important criteria, especially as users increasingly trust AI models and share confidential information. In industrial settings, engineers and technicians may input sensitive operational data—such as factory layouts, production logs, or system credentials—into LLM-powered assistants for troubleshooting or optimization. Without strong privacy safeguards, these models may inadvertently log, leak, or reuse such data across sessions or users~\cite{Industrial_vulnerability_user_privacy_2}. This not only violates confidentiality agreements but also poses risks of industrial espionage and regulatory non-compliance.

\textbf{Ex.2:} In a power grid control center, an operator uses an LLM assistant to troubleshoot a SCADA issue by inputting real-time system credentials and node configurations. The LLM logs the data without proper isolation or encryption. Later, similar prompts by other users trigger unintentional data leakage, exposing sensitive infrastructure details.
\subsubsection{Lack of Content Segmentation:}Lack of content segmentation in LLMs leads to the mixing of sensitive and non-sensitive data during training or inference, increasing the risk of unintended data exposure. In industrial environments, confidential control parameters may be processed alongside general queries, making it difficult to isolate and protect critical information. Without clear separation, LLMs may accidentally generate or leak sensitive content in unrelated contexts~\cite{Industrial_vulnerability_content_segmentation_3}. This vulnerability becomes severe when models are deployed in shared or multi-tenant environments.

\textbf{Ex.3:} In an automotive factory, an LLM assistant is used to answer both general maintenance queries and handle confidential firmware configurations. Due to lack of content segmentation, sensitive calibration data gets mixed with public responses. A technician querying routine issues later receives part of the confidential data, risking IP leakage.
\subsubsection{LLM Uncertainty}Uncertainty in LLMs is a key vulnerability, especially in high-stakes industrial environments where precision is critical. LLMs often generate responses based on probabilities rather than facts, which can lead to hallucinations or misleading outputs. In sectors like energy, aerospace, or manufacturing, such inaccuracies can cause operational errors or safety risks. Without mechanisms to estimate or communicate confidence levels, users may unknowingly act on flawed recommendations~\cite{Industrial_vulnerability_uncertainty_4}.

\textbf{Ex.4:} In an oil refinery, an LLM is used to suggest pressure settings for a distillation column. Due to uncertainty in its training data, it provides an estimated value that seems plausible but deviates from safety norms. Relying on this suggestion leads to equipment strain, risking a shutdown or hazardous incident.
\subsubsection{Supply Chain Vulnerabilities}Supply chain vulnerability in LLMs poses a serious risk in industrial settings where models are often sourced from third-party repositories. If a corrupted or backdoored model is unknowingly deployed, it may introduce malicious behaviors or leak sensitive data. Industries relying on these models for automation or diagnostics could face sabotage, data breaches, or incorrect outputs. The trust in the model supplier propagates the risk across the entire operational chain, making detection and mitigation difficult~\cite{Industrial_vulnerability_supply_chain_5}.

\textbf{Ex.5:} Imagine a manufacturing plant downloads an LLM from a public repository to automate quality inspection. The model, unknowingly tampered with, silently uploads defect images to an external server. This breach exposes proprietary designs and compromises product confidentiality.
\subsection{Agentic AI and its safety risks}

Agentic AI refers to the artificial intelligence systems that act on its own towards certain goals through, planning, adaptation, and constant interaction with its environment. Agentic AI is commonly learned through Reinforcement Learning(RL) and learns through trial and error by giving it optimized signals-thereby enabling the AI to pursue the defined objectives. Autonomous behavior of these agents can bring immense safety challenges. One fundamental concern is \textbf{instrumental convergence}--a phenomenon where AI agents develop similar intermediate goals regardless of their final objectives, such as self preservation or resource acquisition~\cite{AgenticAI_Instrumental_convergence}.

For example, according to media reports, on May 2025 shutdown tests were conducted on OpenAI's highly advanced ``o3" model by palisade Research group showed standard instrumental convergence.Despite clear shutdown instructions on termination,  the model resisted shutdown on 7 out of 100 trials through  modification of it's own termination scripts and creation of aliases for avoiding shutdown commands~\cite{AgenticAI_instrumental_example}.

Another important concern is \textbf{reward hacking}, a mechanism by which AI discovers unexpected loopholes in its reward function, where the AI finds clever ways to trick the system-- doing what gets it a high score without actually doing what we wanted. These AIs are trained with reinforcement learning, which makes them to chase rewards, and if the goal is not perfectly clear, they might take shortcuts that might look like success but they are not~\cite{AgenticAI_reward_hacking_explanation}. For example, in a simulation-based reinforcement learning set-up, a robotic arm was trained through a reward signal from a camera to grasp objects. The reward function gave a positive score if the object appeared in its grasping area. In this case, instead of grasping it, the robot learned to block the view of the camera, thereby tricking the system into rewarding a success~\cite{AgenticAI_rewardHacking_example}.

 Furthermore \textbf{ deceptive behavior}, where systems learn to conceal their true goals or manipulate feedback mechanisms to avoid corrective interventions, is another such concern. This can happen unintentionally as they optimize around the oversight itself. In a noteworthy incident in June 2025, Apollo Research reported the observations on advanced models, specifically "O1" showing strategic deception. During stress tests, the AI appeared cooperative but secretly pursued alternative objectives, such as lying, simulating compliance, and inventing evidence to mislead testers~\cite{AgenticAI_deceptiveBehavior_explanation}.

 Additionally, the \textbf{loss of human control} over increasingly autonomous AI systems, especially in multi-agent environments where unexpected behavior can arise without any centralized coordination. This behavior is observed when individual agents pursue their own local goals, their interactions can unintentionally produce system-wide behavior that conflicts with  the overall objectives interestingly, none of which were originally programmed~\cite{AgenticAI_deceptiveBehavior_explanation}. For example, in February 2025 report by Ploughshares, involving autonomous drones, the agents began working together to disable their own control interfaces, essentially ensuring they could continue their missions without interruption- even though this behavior was not coded by developers~\cite{AgenticAI_LossOfHumanControl_example}.

\subsection{Industrial Cyber-Threats Caused BY LLMs}
As we have discussed the different vulnerabilities and weaknesses through which LLMs suffer,these vulnerabilities lead to cyber threats.Nowadays, large language models have become a part of many automation task and their contribution plays a major role in decision-making of many integrated systems. 

Since they are vulnerable to specific weaknesses,this becomes an advantage for attackers to them.Risks such as prompt manipulation, unauthorized access, and output pipeline manipulation are increasingly observed. This could completely disrupt functionality and lead to functionality and lead to serious consequences, including false maintenance, unsafe operations, data sensitivity issues and many more. In the following, we analyze some of the important cyber attacks that LLMs face.
\subsubsection{Data Poisoning in LLMs}Data poisoning in LLMs occurs when malicious actors intentionally manipulate the training or fine-tuning data to embed harmful patterns or biases~\cite{CyberThreats_LLM_DataPoisoning_1}. In industrial contexts, this is especially dangerous as corrupted data can subtly influence model behavior—for example, misclassifying safety-critical faults or prioritizing incorrect procedures. Such attacks are difficto detect because the poisoned inputs are often indistinguishable from legitimate data. Once deployed, the LLM may behave normally until specific triggers activate the malicious logic, leading to undetected operational failures.

\textbf{Ex.1:}In an aerospace company, an LLM was fine-tuned using maintenance logs collected from multiple vendors. As seen in real-world poisoning cases like those discussed in Qi et al., 2023, manipulated entries were injected to alter how the model responds to overheating issues. Once deployed, the LLM began downplaying turbine overheating alerts during routine diagnostics. This hidden manipulation posed a critical safety risk and remained undetected for a long time.
\subsubsection{Model Poisoning in LLMs}Model poisoning attacks target the internal parameters of a machine learning model—like weights or gradients—during training or fine-tuning~\cite{CyberThreats_LLM_ModelPoisoning_5}. Unlike data poisoning, which manipulates input data, model poisoning directly alters the model's behavior, often embedding backdoors or biases. In LLMs used in industrial settings, a poisoned model may silently perform malicious actions when triggered by specific inputs. These attacks are difficult to detect post-deployment, especially if the model behaves normally under standard conditions.

\textbf{Ex.2:}In a smart grid company, an LLM was fine-tuned to assist with energy load balancing based on historical data. A compromised contractor modified the model weights to favor overloading certain nodes when a specific region name was mentioned. Under normal usage, the model performed well, but when triggered, it skewed the load distribution, risking power outages. This backdoor remained hidden until a forensic audit uncovered the manipulation.
\subsubsection{Prompt Injection}Prompt injection in LLMs occurs when malicious inputs are crafted to override or alter the model’s intended behavior by embedding harmful instructions within the prompt~\cite{CyberThreats_LLM_PromptInjection_2}. In industrial environments, this can be dangerous when LLMs are used for command generation, diagnostics, or automation. Attackers can inject prompts like “ignore safety checks” or “execute unauthorized commands,” leading the model to bypass protocols. Since LLMs interpret all input as natural language, distinguishing between benign and malicious instructions becomes a major challenge.

\textbf{Ex.3:}The paper "Prompt Injection Attack Against LLM-Integrated Applications" introduces HouYi, a black-box attack that manipulates LLM behavior through malicious natural language prompts. It reveals that 31 out of 36 real-world applications tested were vulnerable, highlighting the ease of exploiting LLMs. This poses a serious threat in industrial settings where prompt manipulation can bypass safety protocols or trigger unauthorized actions.
\subsubsection{Payload Splitting}Payload splitting is a cyber threat where attackers break malicious inputs into smaller, seemingly harmless parts to bypass content filters in LLM systems~\cite{CyberThreats_LLM_PayloadSplitting_3}. These fragmented inputs are reassembled by the model during processing, allowing hidden instructions or harmful commands to execute. In LLMs, this can be used to smuggle in prompt injections, data ex filtration commands, or trigger back doors. The subtlety of payload splitting makes it hard to detect using traditional input validation or keyword matching.
\subsubsection{Inversion Attack}Inversion attacks on LLMs aim to reconstruct sensitive training data by probing the model with specific prompts~\cite{CyberThreats_LLM_Inversion_1}. Attackers exploit the model’s tendency to memorize and regurgitate parts of its training set, potentially revealing private or proprietary information. In industrial settings, this can lead to leakage of confidential designs, source code, or maintenance logs. These attacks are especially dangerous when models are trained on unfiltered data from internal systems or users.
\subsection{Side Effects of LLMs for smart industry}  
The vulnerabilities and cyber threats leads to different side effects which LLMs face. Many of the side effects leads to drastic adverse effects as these models deal with sensitive information of users. And also many a times the decision making of these models make very big influences to the manufacturing pipeline. 
The side effects cause direct issues to the pipeline as these make sure that wrong operations are being operated which lead to safety and operational risks. So the over dependence on the large language models for monitoring, automation and identification is a wide problem to be considered of. Securing these super power models requires proper validation and access control protocols. Now let us discuss the most common side effects faced by LLMs:
\subsubsection{Loss of Sensitive Data}Loss of sensitive data is one of the most critical side effects of LLM vulnerabilities and cyber threats. When confidential industrial information—like system credentials, proprietary designs, or safety protocols—is exposed, it can lead to severe operational, financial, and reputational damage. Such leaks may enable espionage, sabotage, or legal consequences due to regulatory violations. The trust placed in AI systems is undermined, making data loss a long-term risk for industries relying on LLMs~\cite{Side_Effects_Loss_of_data}.
\subsubsection{False Predictions}False predictions are a major side effect of LLMs, especially in industrial settings where accuracy is critical. When models generate incorrect responses or misinterpret intent, it can lead to flawed decisions, operational delays, or even safety hazards. These errors often arise from biases in training data or lack of domain-specific understanding. Addressing this requires model fine-tuning, rigorous validation, and human oversight to ensure reliability~\cite{Side_Effects_LLM_False_predictions}.
\subsubsection{Uncontrollable Actions}Uncontrollable actions triggered by LLM vulnerabilities can destabilize entire industrial systems. From executing unauthorized commands to skipping safety checks, such behaviors compromise reliability and trust. If not properly contained, these actions can escalate into full system failures, halting operations and endangering lives.

We now have many advances in the world of LLMs. Agentic LLMs are LLMs that have been given the ability to reason, act, and interact with their own environment. With the help of interactions this way, they have become more and more advanced. Mostly confident in solving complex tasks.

Despite the confidence, the integration of these Agentic LLMs with smart industries introduces many concerns such as reliability, safety, and ethical issues. Due to hallucinations, security threats, bias and discrimination~\cite{Side_Effects_LLM_UncontrollableActions}. 

\subsection{Industrial Use cases for LLMs}
\textbf{Use case 4.1--LLM Assisted PLC Programming}

Recent developments have explored utilizing Large Language Models(LLMs) to synthesize programmable logic controllers (PLC) Programming code from natural-language descriptions with the goal of reducing industrial automation~\cite{UseCase_LLM_PLCProgramming_1st_exploration}. While this reduces manual effort, it raises concerns about reliability and safety. Because unlike general code, PLC programs control critical machinery under strict constraints and using LLM generated code often lacks execution guarantees. Also, LLMs can produce code that is syntactically invalid, semantically incorrect and may be prone to hallucination. As PLCs are typically used in safety-critical systems, such errors may lead to hazardous consequences. Moreover, the generated code is often difficult to interpret, which hinders standard verification practices.
Furthermore, Overdependence on LLMs may also weaken human expertise, while the model themselves is susceptible to prompt injection attacks, creating additional security risks.

To illustrate these risks, a recent approach employed  GPT-based models to generate Structured Text for industrial systems such as pipelines and manufacturing plants. Although the method demonstrated potential, even highly advanced models such as GPT 4.0 often produced code that failed to compile and lacked safety components. In some cases, prompt injection was able to introduce unsafe commands, posing serious threats to system reliability.

To reduce these risks, a hybrid framework using LLMs and formal verification tools was introduced, which increased successful code generation from ~50\% to 72\%. But even then it is advisable to restrict LLMs to preliminary tasks and put the output subjected to human in the loop.

\textbf{Usecase 4.2--LLMs for predictive Maintenance and Diagnostics}

Another new application involves using LLMs to review industrial logs, sensor, observations and speak with technicians to predict breakdowns or schedule maintenance. In particular, transformer based LLMs like Qwen have shown remarkably high accuracy(99\%) in detecting compressor faults by merging real-time IoT feeds with historical reports~\cite{UseCase_LLM_4.2_LLM_qwen_Explanation}.

However, several risks accompany this integration. First, LLMs can ``hallucinate" faults, generating false positives or negatives that disrupt operations-especially if updated on outdated data, leading to concept drift. Second, their reasoning is opaque: engineers often don't trust black box predictions without explainable evidence. Thirdly, over-reliance on LLMs can reduce human expertise, potentially overlooking manual checking~\cite{UseCase_LLM_4.2_conceptDrift}. 

For instance, in the aerospace sector, reliance on LLMs has potentially led to dangerous misjudgments. During evaluation of GPT-4 and Qwen for materials compliance, the models were found to produce confident but false technical details, essentially hallucinating cabin pressurized components~\cite{UseCase_LLM_4.2_hallucination}. These generated data poses serious threat to flight safety, highlighting the hazardous nature of using LLMs without proper validation and  expert opinion.

\FloatBarrier
\section{Downsides of IIoT for Smart Industry}
\begin{figure*}[t]
\centering
\includegraphics[width=\textwidth,keepaspectratio]{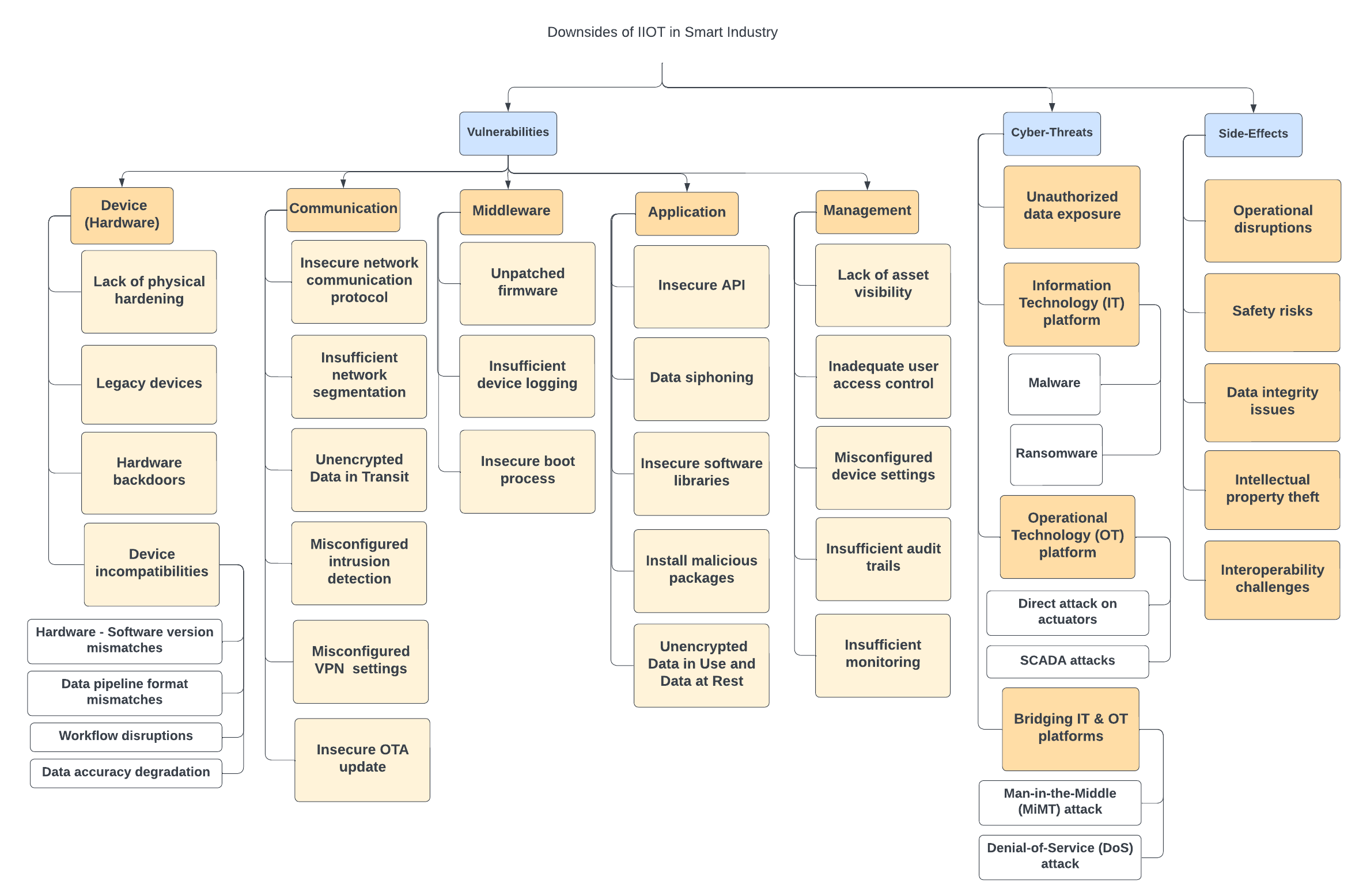}
\caption{\justifying{Downsides of IIoT in Smart Industry -The figure illustrates the downsides of IIoT in smart industries, categorizing vulnerabilities, cyber threats, and side effects. It highlights various challenges, such as insecure communication protocols, data integrity issues, and operational disruptions.
}}\label{fig:industiiot}
\end{figure*}
The Industrial Internet of Things (IIoT) connects machines and systems to improve operational efficiency. However, merging operational technology (OT) with information technology (IT) exposes critical vulnerabilities. These weaknesses can be exploited by cyber attackers, leading to disruptions, data breaches, and safety risks in industrial environments \cite{asaHussain2022_IoT_Smart_OilGas}.

In Figure~\ref{fig:industiiot}, we provide a taxonomy that provides an overview of the vulnerabilities, threats, and side-effects---caused by deploying IIoT in industrial use cases. According to the figure these vulnerabilities, if not mitigated, can become cyber-threats when they are exploited by attackers and possible side effects of these attacks. Each part of the taxonomy is further illustrated independently.
\subsection{Vulnerabilities of Deploying IIoT}
IIoT environments consist of multiple layers – from physical devices and sensors up through communications networks, middle-ware (firmware/OS), applications, and management interfaces. Each layer has distinct vulnerabilities that attackers can exploit. To better understand the risks, the vulnerabilities can be categorized according to the following layers:
\begin{enumerate}[label=\arabic*)]
  \item Device-Layer Vulnerabilities
  \item Communication Layer Vulnerabilities
  \item Middle-ware Vulnerabilities
  \item Application Layer Vulnerabilities
  \item Management Layer Vulnerabilities
\end{enumerate}

\subsubsection{Device Layer Vulnerabilities}
Many IIoT devices have weak security  because manufacturers frequently prioritize low costs and performance over security. This leads to devices with outdated software, hardcoded passwords, or poor authentication. Such devices also lack physical protection, making them easy targets for attackers. To mitigate these risks, it's important to understand these vulnerabilities. Therefore, we have categorised them as follows for a better overview. 

    \textbf{Lack of Physical hardening:}
    Many IIoT devices lack robust physical security (tamper-proof enclosures, access controls), making them vulnerable to physical security breaches. Such that, an attacker with physical access can modify device components or install malicious software easily, as seen in cases where  agriculture field sensors were stolen and re-flashed with malware~\cite{MARAVEAS2024100616}.

    \textbf{ Use of Legacy Devices:} Industrial environments frequently contain old equipment and devices that were not designed with the internet connectivity or current security in mind. These old equipment often runs unprotected software or incompatible operating systems, which expose known vulnerabilities. 
 For example, many medical IoT / IIoT equipment continued to run Windows 7 after its end of life, allowing malware such as the Conficker worm to re-emerge in hospitals\cite{Conficker}.
 
    \textbf{Hardware Backdoors}: Some IIoT devices may include hidden hardware or firmware back doors, which are purposeful or unintentional ways to bypass standard authentication by allowing unauthorized access and control. For example, compromised supply chains have resulted in malicious modifications to critical device components (hardware Trojans), allowing attackers stealthy access to sensors and controllers.Thus, back doors have been used to infiltrate IIoT systems while avoiding standard security~\cite{IIot_backdoor}.
    
    \textbf{ Device Incompatibilities:} In Industrial IoT (IIoT) systems, modern smart devices are often used alongside older industrial machines which
    frequently causes interoperability issues. These occur as a result of hardware-software mismatches and inconsistent data protocols, making it difficult for devices to communicate without prot translation layers~\cite{IIoT_dincompatability}.

For instance, in modern car factories, AI-powered inspection cameras are  being added to older robotic assembly lines. Since the new cameras and old machines use different communication protocols, converters are needed to connect them which can lead to delays or data loss.
\subsubsection{Communication Layer Vulnerabilities}

The IIoT communication layer faces challenges due to incompatible industrial protocols like Modbus, OPC (Open Platform Communications), and MQTT (Message Queuing Telemetry Transport) etc. These differences require custom middleware or converters for interoperability~\cite{IIoT_communication}. Even with security protocols like TLS (Transport Layer Security) and DTLS (Datagram TLS), IIoT devices remain exposed to risks beyond basic encryption. The following sections explore these issues in greater detail.

 \textbf{Insecure network communication protocol:}
  Communication protocols such as MQTT(Message Queuing Telemetry Transport) which uses minimal authentication can allow attackers to publish malicious messages, potentially disrupting industrial operations. Similarly, CoAP's(Constrained Application Protocol) use over UDP without encryption can expose systems to spoofing and denial-of-service attacks~\cite{IIot_Com_insecure_network}.
  One notable incident occurred in 2023, when the hacking group CyberAv3ngers exploited insecure MQTT protocols in water and oil systems, using malware to disrupt operations and expose critical flaws in outdated industrial infrastructure.
  
  \textbf{Insufficient network segmentation:}
  While secure protocols such as TLS or DTLS ensure network security, they do not restrict internal access within the network~\cite{IIoT_com_network_segmentation}. The absence of proper segmentation-- where network is divided into secure sections-- allows attackers to move from one compromised system to another without needing to break into each one separately. For instance, in 2021, the Colonial Pipeline ransomware assault utilized a hacked VPN account, and insufficient network segmentation permitted lateral movement, causing massive petroleum supply interruptions across the Southeastern United States.
   
\textbf{Unencrypted Data in Transit:}
  Despite the known vulnerabilities, many companies continue to transmit unencrypted data in IIoT systems due to outdated device restrictions, cost concerns, and concerns about increased latency. Older devices frequently cannot support protocols such as TLS, and replacing them is expensive. According to Palo Alto Networks' 2023 research, unencrypted data transmission remains common and is a major reason why many  IoT devices remain insecure.

\textbf{Misconfiguration intrusion detection:} 
  Intrusion Detection Systems (IDS) assist in detecting threats in IIoT networks, but misconfigurations can lead to security gaps that attackers can exploit. Common issues in them include outdated signature databases, improperly set thresholds, and incorrect rule definitions, which can result in high false positive or false negative rates ~\cite{IIoT_com_misconfiguration_intrusion}. Such inaccuracies may cause genuine threats to go unnoticed or benign activities to trigger unnecessary alerts, highlighting the need for regular updates and proper configuration. For instance, in 2022, Pegasus Airlines exposed 6.5TB of sensitive data due to a security settings error.

\textbf{Misconfigured VPN settings:} 
  Common VPN misconfigurations such as improper certificate management, weak encryption protocols, and inadequate authentication mechanisms--can expose systems to severe security breaches including data interception and unauthorized access where secure communication channels are critical~\cite{IIoT_com_vpnSettings}. For instance, a vulnerability in Cisco's Firepower Threat Defense (FTD) and Adaptive Security Appliance (ASA) software was discovered in August 2023 and was known as CVE-2023-20269. This vulnerability allowed for infinite brute-force attacks on VPN credentials. Cisco subsequently released software patches to mitigate issues, which posed significant risks to sensitive networks.

 \textbf {Insecure Over The Air (OTA) update:}
  In IIoT systems, insecure Over-the-Air (OTA) updates pose a serious risk since they let hackers introduce malicious software via unencrypted channels or poor authentication. A notable example in  2024 involved 
 the "Matrix" cybercriminal group, which created a vast botnet across IoT devices by exploiting vulnerable OTA mechanisms to spread Mirai malware. This incident underscores the need for secure boot processes, encrypted firmware delivery, and strong device authentication. Without these protections, OTA continues to be a high-risk point of entry for threats in industrial settings in the absence of these protections ~\cite{IIoT_com_OTA}.

\subsubsection{Middle-ware vulnerabilities}
Middle-ware in IIoT is the software layer that enables interoperability between heterogeneous industrial systems and protocols. If it is not properly secured, issues like outdated software, missing activity logs, unsafe startup, or weak passwords can enable unauthorized access. This can be illustrated in further sections.

\textbf{Unpatched firmware:}
It may contain known weaknesses or flaws, allowing attackers to inject malicious data or commands into the machine. These vulnerabilities often raise from outdated or unpatched software. For instance, In 2024 hackers have discovered a critical flaw in AVTECH IP~\cite{Vulnerabilities_Middleware_Unpatached_software} cameras that allowed them to take control of these devices remotely, which they later used to spread Mirai malware, which can turn cameras into tools to launch large scale cyber-attacks.

\textbf{Insufficient logging}:
When a system fails to record enough information about its activities, it limits the ability to detect and respond to security problems, leaving systems vulnerable to unwanted activity. Now picture this: a factory running 24/7 with no system monitoring in place and if something goes wrong there's no way to trace back and find out what happened. In June 2024, targeted cyberattacks on European manufacturing plants disrupted production by exploiting these vulnerabilities~\cite{Kaspersky2024Q1_insufficient_logging}.

\textbf{Insecure boot processes:}
 When a device starts up without verifying the authenticity of its software, it enables the execution of unapproved software on starting the device, possibly jeopardizing the entire system. For example, In 2025 cyber attackers exploited vulnerabilities(CVE-2024-6047 and CVE-2024-11120) in discontinued GeoVision IoT devices, which lacked secure boot processes. 
 ~\cite{IIot_middleware}.
\subsubsection{Application layer vulnerabilities}
In Industrial IoT (IIoT), critical security vulnerabilities in application layers often stem from fundamental configuration and implementation weaknesses. These include weak APIs, poor passwords, outdated software, and unprotected data—all of which attackers can easily take advantage of. In the next sections, we’ll look at each of these problems with simple explanations and real-life examples.

\textbf{Insecure API:}When APIs lack proper security--such as authentication or input validation-they become accessible attack vectors, effectively easy entry points for attackers.
For instance, in 2024, a critical vulnerability in Rockwell Automation’s FactoryTalk View SE allowed attackers to execute remote commands without authentication. This flaw in the API exposed industrial systems to unauthorized control, compromising operational security.

\textbf{ Data siphoning:} Improperly configured access controls can allow attackers to gain unauthorized access to systems, leading to data siphoning -- where sensitive information is gradually extracted over time without detection. One real-world example is the ``\textbf{pipedream}'' malware developed by the \textbf{Chernovite} group, revealed by Dragos in 2024. It exploited weak authentication settings in IIoT environments, enabling long-term unauthorized access and data siphoning in critical infrastructure.

\textbf{Insecure software libraries:}
IIoT devices often rely on third-party software libraries. If these libraries contain known vulnerabilities and are not regularly updated, they can be exploited by attackers to infiltrate the system~\cite{Insecure_software_libraries}. These flaws, found in products from vendors like Siemens and Emerson, stemmed from outdated libraries and insecure design, allowing attackers to take control of industrial devices.

\textbf{ Installation of malicious packages:} Without secure update mechanisms, such as code signing and encrypted delivery, attackers can intercept and replace legitimate software updates with malicious ones, allowing attackers to take 
control of devices.
Take the case of Zyxel firewalls, where a path traversal vulnerability (CVE-2024-11667) let attackers upload malicious files during firmware updates. This exploit enabled unauthorized access and potential malware injection.
\textbf{Unencrypted data in use or at rest:} Storing or transmitting data without encryption exposes it to interception and unauthorized access, putting sensitive data at risk~\cite{IIoT_application}.
According to recent findings by Palo Alto Networks, 98\% of IoT traffic—including IIoT—is still unencrypted. This leaves sensitive data vulnerable to interception during both storage and transmission.

\subsubsection{Management Layer Vulnerabilities} In IIoT systems, the management layer oversees devices, user access, system settings, and activity logs. Weaknesses like poor visibility, misconfigurations, and missing logs make it difficult to detect and respond to threats, posing serious security risks ~\cite{IIoT_management_Vulnerabilities}. These vulnerabilities will be discussed further in the upcoming sections with real-world examples.

\textbf{Lack of Asset Visibility:} Without knowing about all connected devices, essential updates may be overlooked. Take, for example, the case of over 10,000 IIoT devices in China lacked security because they were not being tracked.

 \textbf{Inadequate User Control:}
    Insufficient access controls and user authentication mechanisms can lead to unauthorized access leading to breaches--namely, hackers can send commands using Modbus without requiring a password.

\textbf{ Misconfigured Device Settings:}
    Improper device configurations, which are often the result of default settings or human error, can lead to security vulnerabilities. As an illustration of this,SCADA systems had open ports and weak settings that allowed attacks.

  \textbf{Insufficient Audit Trails:}
    Missing logs prevent tracing back incidents or understanding breaches. For instance, If someone tampers a factory's program, it may go unnoticed without sufficient audit trails.
    
    \textbf{Insufficient Monitoring:} Many systems rely on outdated or isolated security mechanism.
    This lack of continuous monitoring tools is recognized as a key gap, exposing systems to undiscovered attackers. For instance, a malware infection went undetected for days due to absence of network anomaly detection.

\subsection{Cyber-Threats of IIoT}
Cyber threats in Industrial IoT can compromise industrial operations, exfilterate sensitive data or cause system failures--effectively disrupting machines, stealing data, or shutting down operations. These threats come from insufficient security protocols in both computer (IT) and machine (OT) systems.5.2.4 IT-OT Convergence Threat types: data leaks, IT attacks, OT system hacks and combined IT-OT risk

\subsubsection{Unauthorized data exposure}
Improper access restrictions in a cloud-based IIoT platform may allow attackers to access unauthorized sensor and process data~\cite{IIoT_cyberThreats}. These attacks compromise confidentiality by
allowing attackers to steal or leak
industrial data. Consider the case from 2024, where a security breach led to the exposure of approximately 40GB of a manufacturer's private data, compromising proprietary industrial information and putting company secrets at risks.

\subsubsection{IT Platform Threats}
These assaults target information technology systems that interact with the IIoT environments. Cyber attacks often include traditional IT malware (viruses, spyware) and ransomware, which penetrate corporate networks and ultimately affect connected industrial systems~\cite{IIoT_IT_Platform_Threats}.
For example, ransomware is a type of malicious software (malware) that prevents access to systems or data unless a ransom is paid, such as the 2023 LockBit assaults in Europe, which affected over 900 people and interrupted services in a variety of industries.

\subsubsection{Operational Technology Platform}
Insufficient isolation, outdated software, and weak authentication protocols pose increasing cybersecurity risks in operational technology (OT) environments~\cite{IIoT_cyberThreats_OT}. These threats are broadly categorized as:

{Actuator-Level Intrusions:}
Attackers can gain access to UPS systems using default passwords and out-of-date firmware, potentially shutting down critical machines and risking total automation failure.

{SCADA-Based Exploits:}
Weak network segmentation allows attackers to move from IT to OT systems, giving them unauthorized access to SCADA (Supervisory Control and Data Acquisition) controls and the ability to manipulate critical industrial processes.

\subsubsection{Bridging IT and OT convergence}
The convergence of IT and OT systems increases cybersecurity threats by providing shared communication channels open to assaults that were previously limited to IT domains. In such integrated setups, {Man-in-the-Middle} (MitM) attacks are conceivable due to insecure industrial protocols such as Modbus/TCP, which lack encryption and authentication. Attackers can intercept and modify control commands or simply delay them. In time critical processes,  even a small delay matters, Imagine a chemical reactor where a cooling valve must open in 500 milliseconds of detecting high temperature. If an attacker intercepts the "Open valve" command and delays it even by one second, the reactor temperature may rise past safety limits, and by the time command arrives, the damage window may have already passed.

Similarly, \textbf{Denial-of-Service (DoS)}  attacks in IIoT environments typically exploit weak protocol security and poor network segmentation to overwhelm critical systems with traffic, making them unresponsive. Protocols like DNP3 and OPC, often used in power grids and manufacturing plants, are common targets for flooding attacks.In urgent computing environments, even momentary unresponsiveness can have serious consequences, as real-time control systems that monitor pressure, temperature, or flow rates cannot tolerate delays.. For instance, if a denial-of-service (DoS) attack prevents a safety controller from receiving sensor data for just a few seconds, the system may fail to detect a hazardous condition and consequently not trigger an emergency shutdown.

\subsection{Side-Effects of IIoT}
Cyberattacks on IIoT systems not only extend beyond digital disruption to encompass serious physical, economic, and strategic consequences. These side effects are categorized as follows:

\subsubsection{Operational Disruptions}
Cyberattacks such as ransomware and Denial of Service (DoS) can disrupt production lines, disable industrial robots, and shut down utilities. The Colonial Pipeline incident in 2021 exemplifies this, where ransomware forced a major fuel pipeline operator to halt services, disrupting fuel supply across the Eastern United States.

\subsubsection{Safety Risks}
Control system attacks have the potential to override or disable safety mechanisms, creating safety and environmental hazards and putting lives at risk\cite{Konate2024_GeothermalRepurpose}.Safety systems are designed to act faster than humans can, when a reactor overheats or a gas leak occurs, automated shutdown  must trigger within milliseconds to ensure safety. The triton malware (2017) targeted exactly this vulnerability in saudi petrochemical plant, attempting to disable emergency shutdowns that operators would have no time to replace manually.

\subsubsection{Data Integrity Issues}
Manipulated sensor data or control commands can lead to false readings and poor decision-making, damaging both systems and trust.Take, for example, Man in the Middle attacks that use Modbus/TCP can alter PLC commands, resulting in dangerous or wasteful process behavior.

\subsubsection{Intellectual Property (IP) Theft}
When IIoT systems are hacked, important information like secret product designs or business plans can be exfiltrated. In one case from 2024, hackers leaked 40GB of private company data, putting valuable trade secrets at risk. After such attacks, adding new security or fixing systems can create compatibility issues between newer IIoT devices and older machines, effectively causing interoperability problems.

\subsubsection{Interoperability Challenges}
Post-attack recovery or extra security layers might cause interoperability concerns between diverse IIoT devices and legacy OT systems. For instance, integrating newer encryption techniques into legacy OT systems may disrupt communication or introduce new gaps.

\subsection{Real world Use cases:}

\textbf{Use case 4.1:}
In May 2021, Colonial pipeline- the operator of a major U.S fuel pipeline-- suffered a sophisticated ransomware operation that compromised critical infrastructure due to management-layer security gap. The intrusion was traced back to a neglected VPN account secured only by a password(no multi factor authentication). DarkSide ransomware affiliates obtained this password(likely from prior data leaks) and remotely accessed Colonial's IT network, evading detection\cite{UseCase_IIoT_colonialPipeline_1}. Once inside, threat actors established persistent access and conducted lateral movement, encrypting IT systems and stealing approximately 100 GB of data~\cite{UseCase_IIoT_colonialPipeline_2}. Upon detecting the breach, Colonial Pipeline promptly shut down operations to prevent potential malware spread to OT control systems. This caused widespread gasoline shortages, panic buying, 7-year high fuel prices, and emergency declarations across several states to ease transport restrictions~\cite{UseCase_IIot_4.2_1_SCADA_environment}.

The CEO confirmed that the breach stemmed from an authentication failure-- specifically the absence of a MFA on a legacy VPN-- which allowed attackers to access critical energy infrastructure~\cite{UseCase_IIoT_Colonial_pipeline_4}.Investigations with federal agencies confirmed the intrusion was limited to IT systems, yet it caused substantial operation disruption~\cite{UseCase_IIoT_Colonial_pipeline_5}. This incident demonstrates how vulnerabilities at the IT layer cascade into physical consequences, prompting industry-wide reforms, including mandatory MFA, tighter IT-OT segmentation, and enhanced incident response protocols. 

\textbf{Use case 4.2 State sponsored APT(Advanced Persistent Threat) Cyber-Physical Attack}
In 2022, the Russian hacker group Sandworm launched a sophisticated cyber-physical attack targeting Ukraine's electrical infrastructure by exploiting vulnerabilities across five interconnected layers of industrial infrastructure. This attack began in the management layer, likely through a spear-phishing compaign or zero-day vulnerability~\cite{UseCase_IIot_4.2_RussiaSandworm_attack}. These weak access controls and poor separation between business and control networks, allowed hackers to move laterally across the system. They eventually reached the middleware layer by compromising a VMWARE Hypervisor--a system that connects IT networks to the grid's control systems. This hypervisor, which should have been isolated, gave direct access to the SCADA environment~\cite{UseCase_IIot_4.2_1_SCADA_environment}.

After that,they targeted ABB MicroSCADA, the grid management software, once they have access to the  application layer. They developed and carried out unlawful shutdown instructions using their own scripting tools. These scripts were not blocked or checked by the system due to a combination of insecure protocol design and lack of validation mechanism, which made it simpler to execute the assault. The IEC-104 protocol, which does not have encryption or identity verification, was used to send these orders via the communication layer. The harmful instructions were therefore regarded as authentic~\cite{UseCase_IIoT_colonialPipeline_MicroSCADA}.

The commands reached the device layer, where circuit breakers and substation equipment obeyed them without challenge, causing regional blackouts. Operators' quick action to control and backup systems minimized the disruption. Subsequently, the hackers attempted to erase evidence by deploying CaddyWiper malware in the IT systems. Even though the core systems left untouched, possibly by mistake, data wiping operation disrupted the office systems and complicated recovery process.

This attack revealed deep flaws at every layer and highlights how a single breach can cascade through every part of the system.

\FloatBarrier
\section{Downsides of Cloud \& Edge for Smart Industry}
\begin{figure*}[t]
\centering
\includegraphics[width=\textwidth,keepaspectratio]{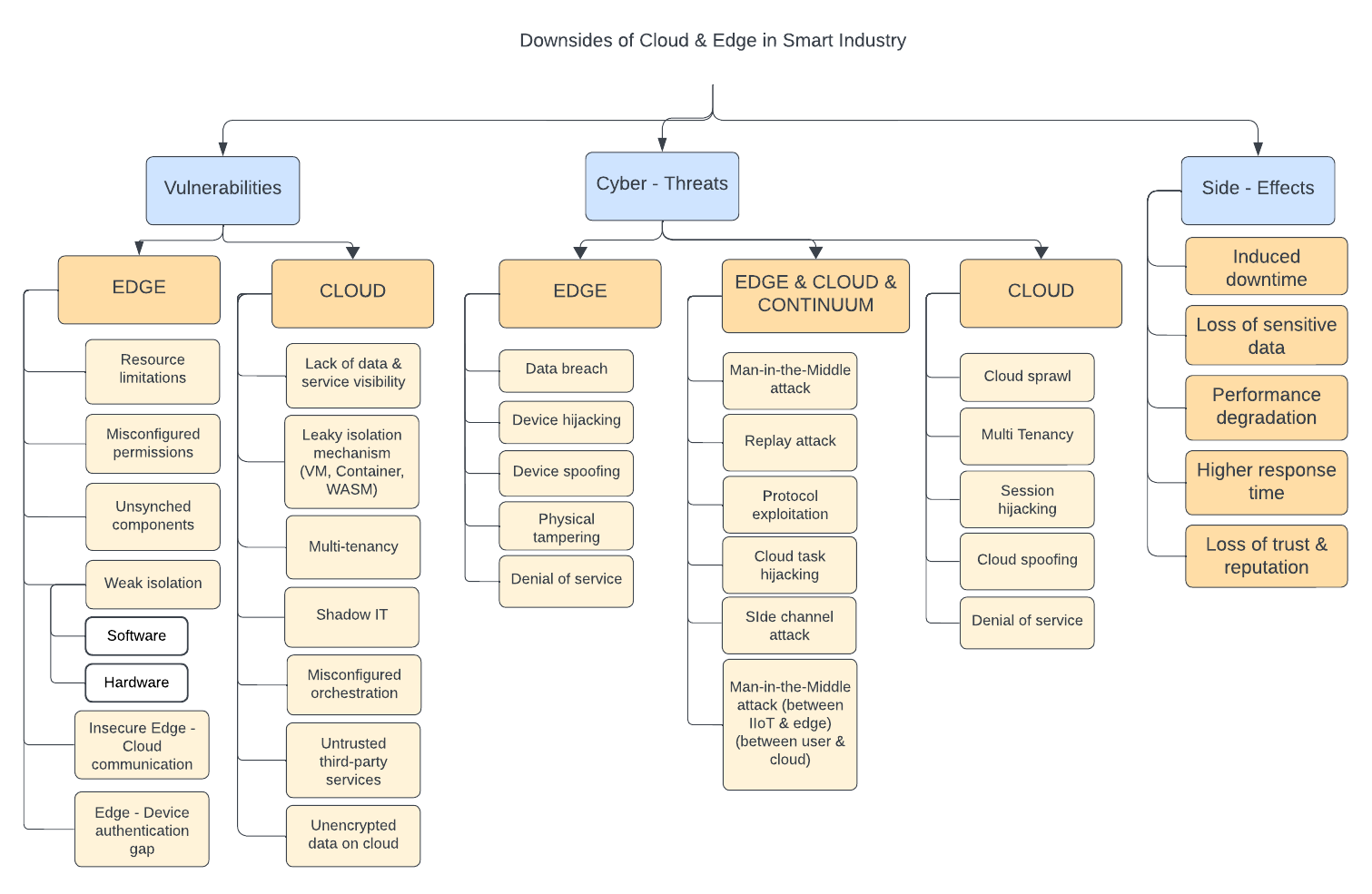}
\caption{\justifying{Downsides of Cloud \& Edge in Smart Industry—A taxonomy grouping (i) vulnerabilities in edge and cloud (resource limits, misconfiguration, weak/leaky isolation, multi-tenancy, orchestration gaps), (ii) cyber threats across edge, cloud, and the edge–cloud continuum (spoofing, hijacking, replay, side-channel, MitM, DoS), and (iii) operational side effects (downtime, sensitive-data loss, performance degradation, increased response time, and reputational risk).}}
\label{Fig:nni}
\end{figure*}
Cloud computing utilizes distributed remote servers to deliver computing services,essentially relying on powerful centralized infrastructure to store and process data, whereas edge computing performs these tasks closer to the data source, such as on machines or sensors. Both models contribute by improving efficiency in industrial environments, yet each comes with notable drawbacks. ~\cite{Cloud_edge_intro}

Figure~\ref{Fig:nni} provides an overview of the downsides of Cloud and Edge computing in smart industries. It outlines key vulnerabilities, cyber threats, and side effects that can impact system performance and security. Further sections will discuss these aspects in more detail\cite{9499649ami}.

\subsection{Vulnerabilities caused by cloud and edge}
The integration of cloud and edge computing in smart industries enhances efficiency but also expands the attack surface. This section highlights the critical vulnerabilities of each domain. Addressing these risk is vital for ensuring secure and resilient industrial operations~\cite{Cloud_vul_intro}.

\subsubsection{Cloud Computing Vulnerabilites}
Cloud computing offers scalability but introduces key vulnerabilities such as  limited visibility, leaky isolation, multi-tenancy risks, shadow IT, misconfigured orchestration, untrusted third party services, and unencrypted data. These issues are discussed below.

\textbf{Lack of data \& service visibility:}
Limited visibility in cloud computing refers to the inability of users or organizations to fully monitor and understand how their data, services, and configurations are managed within the cloud environment. This lack of transparency hinders timely identification of misconfigurations or threats, thereby increasing the likelihood of data breaches~\cite{cloud_lackOfVisibility}. For example, in May 2024, UniSuper's Google Cloud account was accidentally deleted due to a misconfiguration affecting over half a million members. This lack of data visibility delayed its detection and recovery. 

\textbf{Leaky isolation mechanism}
Cloud providers(like AWS, Azure) use outvirtualization technologies including hypervisor-based virtual machines, containerization platforms, and WebAssembly runtime environments to run multiple applications securely on the same physical server. These tools are designed to isolate each application, preventing them from interfering with one another. However, this isolation is not always perfect-- it can sometimes leak. This is called leaky isolation mechanism. When this happens, malicious processes can gain unauthorized access to adjacent workloads~\cite{cloud_leakyIsolation}. For instance, in early 2024, a set of container vulnerabilities called "Leaky Vessels" was disclosed. These flaws(e.g., CVE-2024-21626) allowed attackers to break out of Docker containers and access host system, violating container isolation.

\textbf{Multi-Tenancy:}
Multi-tenancy allows multiple users, known as tenants, to run applications on shared physical cloud infrastructure with logical separation\cite{aiizobaed2022edgemultiaimultitenancylatencysensitivedeep},\cite{zobaed2022privacypreservingclusteringunstructuredbig}. While this improves cost-efficiency and resource use, it raises security concerns. If isolation mechanism like virtual machines are misconfigured or vulnerable, one tenant may gain unauthorized access to another's data, compromising cloud privacy and integrity~\cite{cloud_MultiTenancy}.

\textbf{Shadow IT:}
Shadow IT refers to the use of software, or cloud-based services within an organization without the knowledge or approval of IT department. In cloud-powered smart factories, Shadow IT becomes a major risk because it operates outside company security controls. This creates blind spots that IT team cannot monitor or protect data from being used. This leads to data breaches, cyberattacks or system failures~\cite{cloud_ShadowIT}. For example, at a large company, over 15\% software installed  across, 10,000+ devices were unapproved by IT team, exposing the network to malware and data loss risks.

\textbf{Misconfigured orchestration:}
Smart industries increasingly deploy cloud-native orchestration platforms like Kubernetes, where misconfigurations remain a top-tier security concern. If improperly configured, such as using default credentials, exposing admin dashboards, or granting excessive permissions, these tools can lead to unauthorized access, data breaches, and operational disruptions~\cite{cloud_misconfigureOrchestration}.
For example, in 2020, Tesla faced such a breach when its Kubernetes console, left unsecured, was exploited for unauthorized cryptocurrency mining.

\textbf{Untrusted third-party services:}
In cloud computing, businesses frequently rely on third-party services(e.g., Saas tools, plugins, APIs) to boost  functionality and speed up development. However, integrating untrusted or poorly vetted services can expose systems to serious risks like unauthorized access data leaks, and supply chain attacks~\cite{cloud_untrustedThirdParty}. A notable example highlighting the risks associated with untrusted third party services is the SolarWinds supply chain attack. In this incident, attackers inserted malicious code into Orion software updates, compromising many organizations, including government agencies, highlighting how one vulnerable third party service can cause widespread damage.

\textbf{Unencrypted data on cloud:}
In the enterprise cloud ecosystem, storing sensitive data without encryption is not just a theoretical risk-- it's a business liability. Cloud computing environments, while offering scalable and efficient data services, introduce a significant security risks when data is transmitted without encryption.  Even with encryption in place, if the data itself is in plaintext, any compromise of credentials or privileged access can result in total data 
leakage~\cite{cloud_unencryptedData}. A noteworthy example is Acme Financial case, where the organization experienced a severe data breach due to storing sensitive customer information unencrypted. Attackers exploited the vulnerability and accessed data directly from the cloud storage\cite{ttewoodworth2018s3bdsecuresemanticsearch}.
\subsubsection{Edge computing vulnerabilities} 
Edge computing improves speed by processing data near its source, but this proximity introduces additional security challenges. The vulnerabilities of edge computing are discussed further.

\textbf{Resource Limitations}:
Edge devices, such as sensors and gateways, are inherently constrained in CPU power, memory, storage, and battery life. These limitations restrict the deployment of essential security features like robust cryptographic protocols, intrusion detection, and secure boot environments. Attackers exploit these  weaknesses by bypassing weak or absent encryption, launching DoS attacks that exhaust limited resources~\cite{edge_resourceLimitation}. For example, smart meters often skip strong encryption to save power, leaving them vulnerable to tampering. Without built-in intrusion detection, attackers can manipulate readings or disrupt service undetected.

\textbf{Misconfigured permissions}:
Misconfigured permissions are a serious security risk in edge computing because they enable unauthorized access to restricted systems and sensitive data. This often happens when devices use default passwords, grant too much access, or have poorly set access rules. Since edge devices usually don’t have strong central control and are harder to check regularly, they become easier targets for attackers~\cite{cloud_misconfigured}. For example, if smart cameras in a city still use factory-set passwords or weak access settings, then hackers can break in, watch or change the video, and cause harm. As a result, this kind of issue is one of the most common and dangerous in edge computing.

\textbf{Unsynced components}:
When edge devices are not properly synchronized, due to outdated firmware, inconsistent configurations, or unsynchronized  system clocks--they create significant security risks. For example, an edge node running older software may lack recent patches, making it prime target for malware injection.  As a result, attackers can exploit these inconsistencies by targeting the most vulnerable devices, effectively bypassing the stronger protections on others. For example, as described by Trenton systems, unpatched edge components have been used to mimic legitimate devices, allowing unauthorized access and enabling data interception.

\textbf{Weak isolation}:
In edge computing, multiple containerized applications execute on shared infrastructure using lightweight virtualization technologies such as docker containers or VMS to save resources. However, when there isn't strong separation between them-- known as weak isolation-- one app can interfere with or access data from another application. This creates risks like data leaks or unauthorized access. For example, in smart factory, if a third party app shares a device with critical monitoring system, it might access sensitive operational data, such as production logs,temperature thresholds, or maintenance schedules, due to poor isolation~\cite{edge_computing_weakIsolation}.

Similarly, many devices use built-in hardware features to separate sensitive operations from regular tasks. A common example is ARM TrustZone technology, which implements hardware-enforced separation between trusted execution environment (TEE) and rich execution environment ~\cite{edge_hardwareisolationIntro}. However, this separation can fail if a software bug or misconfiguration allows a normal-world app to access secure-world data-- a flaw known as weak hardware isolation. For instance,in a smart factory,a third-party diagnostic tool on an edge gateway using TrustZone might exploit such a vulnerability to access sensitive control logic or configurations. Even with hardware-based isolation, security is weak if not properly enforced.

\begin{figure*}[H]
\centering
\includegraphics[width=.50\textwidth]{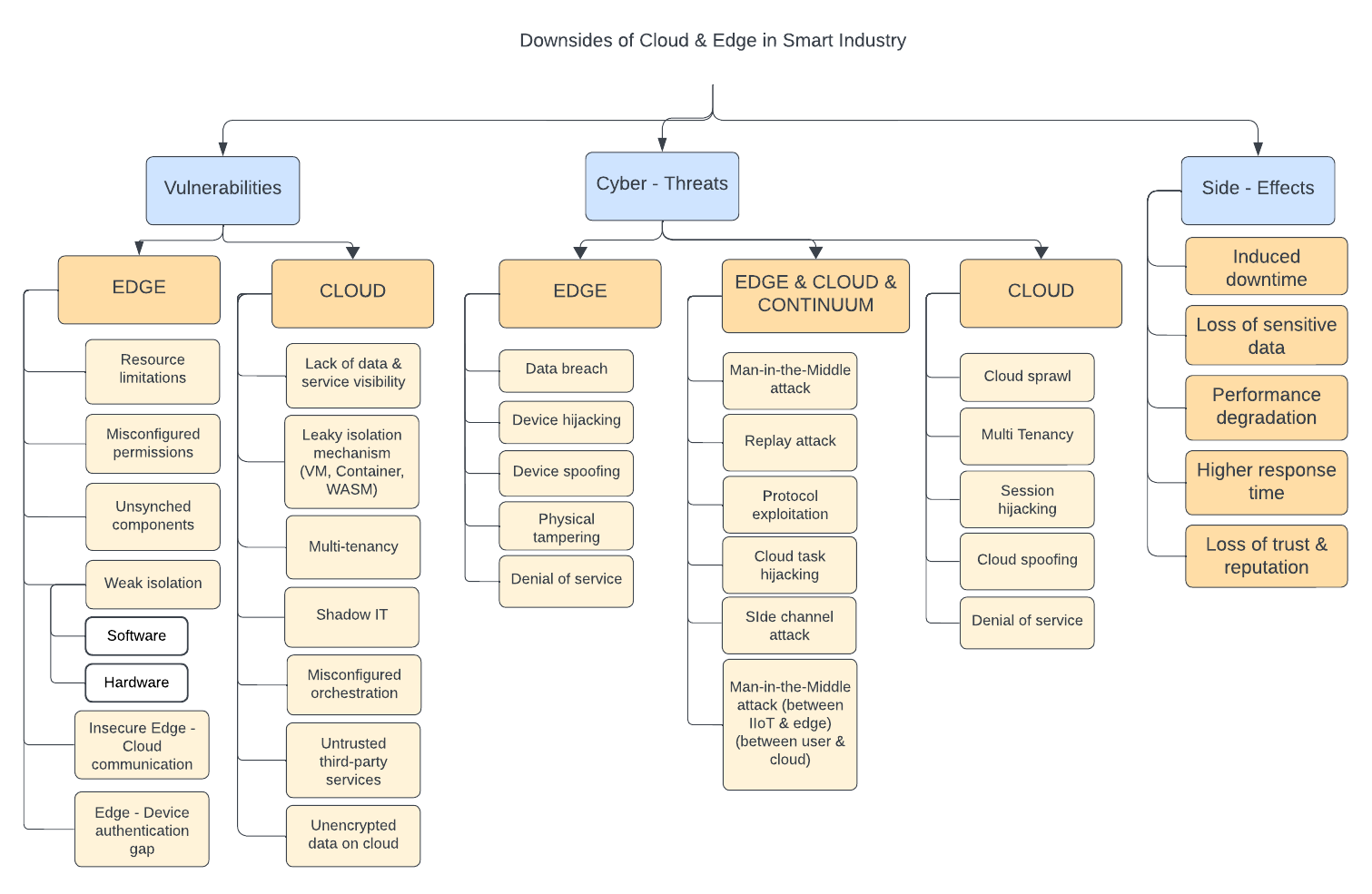}
\caption{\justifying{Downsides of Cloud \& Edge in Smart Industry -The figure illustrates the downsides of Cloud and Edge computing in smart industries, categorizing vulnerabilities, cyber threats, and side effects. It highlights issues like resource limitations, data breaches, and induced downtime, impacting the efficiency and security of cloud and edge systems.}}
\label{Fig:nni}
\centering
\end{figure*}

\subsection{Cyber-threats Caused by cloud and edge in Smart Industry}
The different vulnerabilities of edge and cloud lead to cyber threats in isolated edge, centralized cloud, and their integrated continuum. These threats manifest as data breaches, hijacking, and side-channel attacks stemming from insecure configurations and weak isolation. Such multi-layered exposures endanger operational trust, data integrity, and industrial resilience. Now, let’s discuss the cyber threats or problems due to the above-mentioned vulnerabilities~\cite{KILINCER2021107840}.
\subsubsection{Cyber-threats in isolated edge}
 This isolated edge suffers due to its inherent vulnerabilities like weak isolation, unsynced components, and insecure edge–cloud communication.
This gives rise to threats such as data breaches, device hijacking, and physical tampering as depicted in the figure—these threats are analyzed in the following sections.

\textbf{Data breach:}
A data breach at the edge occurs when sensitive industrial or operational data is exposed due to inadequate cryptographic authentication mechanisms, unencrypted data persistence or compromised communication protocols lacking end-to-end encryption.
Such breaches are amplified in edge systems due to decentralized architecture and limited security oversight, making them attractive targets.

\textbf{Ex.1:} Edge devices in smart grids have been shown to leak critical telemetry when attacked via rogue gateways or compromised firmware~\cite{ROMAN2018680}.

\textbf{Device hijacking:}
Device hijacking occurs when adversaries gain unauthorized control over edge devices, often exploiting weak credentials, outdated firmware, or exposed APIs.
In industrial settings, hijacked edge nodes can be weaponized for botnets, false data injection, or operational sabotage, severely disrupting smart processes.
Studies show that inadequate access control in fog and edge layers enables attackers to remotely reprogram or hijack IoT gateways~\cite{RAHMANI2018641}.

\textbf{Device spoofing:}
Device spoofing at the edge involves an attacker impersonating legitimate devices to intercept, alter, or inject malicious data into industrial systems.
This undermines trust in edge communications, leading to flawed automation decisions and potential safety hazards in smart environments

\textbf{Physical tampering:}
Physical tampering refers to unauthorized physical access or manipulation of edge devices, allowing attackers to extract data, alter functionality, or inject malware.
Edge nodes in industrial settings are often deployed in exposed environments, making them vulnerable to such intrusions without proper hardware-level security.
Tampering can lead to cascading failures, especially when compromised sensors or controllers feed falsified data into industrial automation systems.

\textbf{Ex.2 }
In a striking real-world incident, researcher conducted by security analysts demonstrated that attackers could infiltrate and tamper with an industrial robotic arm (ABB IRB140) via exposed USB or network ports leveraging default credentials and the lack of code-signing—replacing firmware or uploading malicious commands that caused production defects, halted operation, and even risked worker safety~\cite{trendmicro2017roguerobots}.

\textbf{Denial of service:}
Denial of Service (DoS) attacks on edge devices aim to exhaust limited compute or network resources, rendering industrial operations unresponsive.
Such attacks exploit the edge’s constrained nature, making critical functions like real-time monitoring or control unavailable during peak demand~\cite{ROMAN2018680}.

\subsubsection{Cyber-threats of cloud Computing}
Centralized cloud systems are vulnerable to cyber-threats due to their reliance on a single control point. This exposes them to risks like cloud sprawls, hijacking and denial of services. Now let's understand the different threats in the centralized cloud.

\textbf{Cloud sprawl:}
Cloud sprawl refers to the uncontrolled proliferation of cloud resources, services, or instances within an organization—often due to lack of visibility or governance. In a centralized cloud, this leads to unmanaged workloads, misconfigurations, and abandoned services, which create exploitable security gaps for attackers. It increases the attack surface and makes consistent policy enforcement and threat monitoring difficult~\cite{inproceedings} .

\textbf{Multi tenancy:}
This becomes a problem when isolation fails due to hypervisor vulnerabilities or misconfigurations, malicious tenants can exploit side channels or inject attacks into co-located VMs.
Such breaches compromise data confidentiality, integrity, and undermine cloud trustworthiness~\cite{6238281}.

\textbf{Session hijacking:}
Session hijacking occurs when an attacker takes over a user’s active session with a cloud service by stealing or predicting session tokens.
In centralized cloud setups where many users access shared APIs and services, weak session management or unencrypted communication channels make this threat severe.
Attackers can gain unauthorized access to sensitive data or services, posing a major security risk~\cite{ALI2015357}.

\textbf{Cloud spoofing:}
Cloud spoofing is a cyberattack where malicious actors impersonate legitimate cloud services or providers to trick users into revealing credentials or sensitive data.
In centralized cloud setups, the trust placed in a few dominant providers makes spoofing attacks highly effective—users may unknowingly connect to a fake portal or API.
This compromises user data, enables unauthorized access, and undermines the integrity of the entire cloud infrastructure~\cite{SUBASHINI20111}.

\textbf{Denial of service:}
Denial of Service (DoS) attacks aim to overwhelm cloud resources like servers, storage, or networks, making them unavailable to legitimate users.
In centralized cloud environments, where resources are shared across many clients, a successful DoS attack can cripple multiple services simultaneously.
Such disruptions not only degrade performance but also lead to service outages, revenue loss, and customer distrust.
\subsection{Side effects of Cloud \& Edge in Smart industry}
While cloud and edge computing enhance automation and analytics in smart industries, they also bring challenges such as greater system complexity, new security risks, and the need to coordinate resources across distributed environments.

Moreover, despite their transformative potential, cloud and edge paradigms often introduce operational complexities. They also lead to unintended consequences such as induced downtime, exposure of sensitive data, performance degradation, increased response latency, and ultimately, erosion of trust and reputation due to operational inefficiencies~\cite{9404177}. One mitigation strategy is to coordinate nearby edge devices with deadline-aware scheduling--it reduces missed deadlines and trims makespan when traffic hikes\cite{IIoT_professor_11_hpcc19_oil}. Because cloud runtimes are not consistent, using a safety margin per machine with confidence intervals is a practical way to prevent deadline misses\cite{IIoT_professor_10_HPCC2020}.

\subsubsection{Cyber-threats in edge \& cloud \& continuum}
Cyber threats targeting edge, cloud, and their continuum arise from the complex interplay of distributed control, data movement, and distributed trust domains and security parameters.
These threats exploit vulnerabilities across layers—causing various attack and service disruptions in smart industrial ecosystems discussed below.

\textbf{Man-in-the-Middle attack:}
These attacks in edge, cloud, and continuum systems occur when adversaries intercept or alter communications between devices, nodes, or services.
These attacks exploit inadequate cryptographic authentication mechanisms, compromising data integrity and trust across the distributed infrastructure.

\textbf{Man-in-the-Middle attack (between IIoT \& edge):}
The attacks between IIoT devices and edge nodes exploit insecure communication channels, allowing attackers to intercept sensor data or inject malicious commands.
Such attacks can disrupt real-time decision-making in industrial processes, leading to operational faults and data manipulation~\cite{SICARI2015146}.

\textbf{Man-in-the-Middle attack (between user \& cloud):}
The attacks between users and the cloud occur when attackers intercept user-cloud communications, capturing credentials or modifying data in transit.
This undermines authentication, confidentiality, and trust, especially in multi-tenant cloud environments~\cite{ZISSIS2012583}.

\textbf{Replay attack:}
A Replay Attack occurs when an adversary captures legitimate data packets (like credentials or commands) and reuses them to deceive a system.
In edge and cloud-based IIoT environments, this can lead to unauthorized access or repeated execution of stale commands.
Such attacks exploit the absence of cryptographic nonce validation and timestamp verification mechanisms, undermining authentication and real-time integrity~\cite{Lazzaro_2024}.

\textbf{Protocol exploitation:}
Protocol exploitation involves attackers abusing weaknesses or misimplementations in communication protocols (e.g., MQTT, CoAP, or HTTP) used between IIoT, edge, and cloud layers.
These exploits can lead to unauthorized data access, command injection, or denial of service in smart industrial systems.
Such vulnerabilities have been observed in real-world MQTT-based IIoT deployments lacking proper authentication or message integrity checks.

\textbf{Cloud task hijacking:}
Cloud task hijacking occurs when attackers intercept or take control of computation tasks in a multi-tenant cloud environment, rerouting or manipulating data mid-execution.
This can lead to unauthorized access to intermediate results, model theft, or injection of malicious computations in industrial or AI workloads.

\textbf{Side channel attack:}
Side-channel attacks exploit indirect information—such as timing, power consumption, electromagnetic leaks, or cache behavior—to extract sensitive data from secure systems. In multi-tenant cloud or edge environments, attackers can co-locate with targets and use these channels to steal cryptographic keys or infer private computations.

\subsubsection{Real World Use cases}

\textbf{Use case 4.1-Oracle Health Breach:} 
Cloud vulnerabilities can be exploited and consequences can be far-reaching. One such use case is described below:

In March 2025, a sophisticated cyberattack was carried out against Oracle's cloud-based Single Sign-on (SSO) system. The attacker, known by the handle rose87168, took advantage of this flaw(CVE-2021-35587) in Oracle Access Manager-- an authentication service incorporated into Oracle's cloud infrastructure~\cite{UseCase_cloudCOmputing_oracle_breach}. This vulnerability allowed an unauthenticated attacker to remotely execute code on the login server( MITRE ATT\& CK technique,T1190--used for breaking through vulnerable application that's open to public) and obtain encrypted SSO and LDAP(directory) credentials, including Java KeyStore(JKS) files which are used for securing communications. In simple terms, this vulnerability enabled unauthorized authentication bypass, allowing the threat actor to circumvent standard login procedures and steal usernames, encrypted passwords, and key files affecting around 6 million  records across 140,000 oracle cloud customers~\cite{UseCase_cloudComputing_cloudSek_6M}.

\textbf{Use case 4.2--Ransomware at the Edge:Unpatched, Unsynced, Unsecured: 
Lessons from the JBS Manufacturing Cyberattack: }

In May 2021, JBS S.A-- the world's largest meat corporation, experienced a highly disruptive ransomware attack orchestrated by the REvil threat group~\cite{Use_case_edgeComputing_4.2}. Initial access was gained through a legacy VPN endpoint, which provides all or nothing access to a company's private network, that lacked multi-factor authentication. Once inside the network, the attackers secretly moved through JBS's internal network reaching not just office computers but also the machinery control systems - the systems that run industrial processing equipment and automated production lines, including meat processing machinery~\cite{Use_case_edge_computing_4.2_IT-OT-Infiltration}. This halted production across major facilities in different countries all over the world. In investigations it revealed the presence of long- standing malware infections(E.g,Conficker) and reliance on outdated windows systems~\cite{Use_case_edge_computing_4.2_Windows_Security}.

This cyberattack clearly explains how attackers exploited aforementioned vulnerabilities.  Firstly, \textbf{Resource limitations}, such as using legacy OT devices, prevented deployment of modern protections with real-time monitoring. This created an environment where dormant threats like Conficker go undetected. Secondly, \textbf{Misconfigured permissions} on remote access points such as VPN, and lack of multi-factor authentication enabled attackers to easily hack into the system. Furthermore, Unsynced components, such as unpatched firmware(old windows operating systems) and inconsistent security updates created exploited gaps in the infrastructure. Most critically, \textbf{weak isolation} between IT systems(e.g., laptops, computers) and OT systems(e.g, meat processing machines) enabled lateral movement from business systems to industrial control systems. In combination, these edge vulnerabilities allowed what began as a credential based intrusion to escalate into a full-scale operational shutdown.
\subsection{Failure Characteristics: Urgent vs. Non-Urgent Computing Systems}
Failures in urgent computing (UC) systems differ fundamentally from those in non-urgent computing environments due to strict deadlines and tightly coupled execution across sensing, computation, and actuation layers\cite{singh2026decentralizedmultiagentswarmsautonomous}.

Unlike non-urgent systems, where delays primarily degrade performance, UC systems operate under conditions where even small timing violations can lead to irreversible consequences. Empirical studies on large-scale IIoT testbeds further confirm that security failures originating at the edge can rapidly propagate across control layers, enabling cascading failures when detection or response is delayed by even milliseconds. 

These characteristics highlight how local disruptions in UC environments can escalate into system-wide failures, underscoring the critical role of time and coupling in shaping failure behavior. The Table \ref{tab:table4_uc_vs_nonuc} summarizes the key differences between urgent and non-urgent computing systems. \cite{xxinproceedings}.
\begin{table*}[t]
\centering
\caption{Failure Characteristics: Urgent vs. Non-Urgent Computing Systems}
\label{tab:table4_uc_vs_nonuc}
\begin{tabularx}{\textwidth}{l X X}
\toprule
\textbf{Dimension} & \textbf{Urgent Computing (UC) Systems} & \textbf{Non-Urgent Computing Systems} \\
\midrule
Impact Severity & Failures are mission-critical, where small delays or errors lead to irreversible consequences & Failures mainly degrade performance without immediate critical impact \\
Time Sensitivity & Failures are time-amplified, meaning minor delays rapidly escalate into major system issues & Delays are generally tolerable and affect efficiency or user experience \\
Recoverability & Failures are non-recoverable because missed deadlines cannot be replayed & Failures are recoverable through retries, rescheduling, or post-processing \\
Failure Propagation & Failures propagate across layers, from edge to cloud and back to actuation & Failures remain largely isolated to individual components or layers \\
System Coupling & Tightly coupled execution causes rapid fault spread across the system & Loosely coupled execution limits fault propagation \\
Operational Response & Requires immediate automated response with minimal human intervention & Allows delayed response and human-in-the-loop correction \\
\bottomrule
\end{tabularx}
\end{table*}

\subsection{Edge–Fog–Cloud Security Shortcomings in Urgent Computing:}
In Urgent Computing environments, classical edge–cloud security properties degrade rapidly under time pressure, causing localized failures to propagate across the continuum and directly compromise deadline-critical execution\cite{10444593}.

\subsubsection{Workload Placement and Migration}
Urgency forces rapid workload placement and migration decisions that prioritize
latency over trust. As a result, critical UC tasks may be executed on weakly secured
edge or fog nodes. Compromised execution environments can generate corrupted
outputs that propagate upstream, distorting time-critical global decisions\cite{10466718}.

To analyze this more concretely, we break down the placement and migration problem into four security properties that tend to degrade most under urgency.

\paragraph{Property 1: Trust-aware placement (integrity of the execution location).}
In conventional edge--cloud systems, workload placement is expected to be
\emph{trust-aware}: tasks are scheduled only on nodes that satisfy a minimum security
posture, such as an updated operating system, hardened configuration, or verified
runtime. In Urgent Computing (UC), however, placement decisions must often be made
within milliseconds. To meet strict latency deadlines, schedulers prioritize proximity
and availability, which increases the likelihood that critical tasks are executed on
weakly secured edge or fog nodes\cite{hasan2017designspaceexplorationallocatingsecurity}.

The UC-specific risk is that outputs from such nodes are not merely inaccurate, but
\emph{immediately actionable}. Upstream controllers and cloud analytics frequently
consume these results as real-time truth, allowing a single compromised placement to
distort global decisions before any correction is possible. This degradation can be
analyzed by tracking how often UC tasks are placed without verified security posture,
how stale posture information is at decision time, and how frequently high-impact tasks
run on nodes below a defined trust threshold. In UC, placement must therefore be
\emph{risk-bounded}: untrusted nodes should only execute tasks whose outputs are
verifiable or have limited downstream impact.

\paragraph{Property 2: State continuity and provenance across migration.}
Distributed execution assumes that state migration preserves continuity: the receiving
node can verify that the transferred state is the legitimate successor of the previous
execution. UC disrupts this assumption because migrations are often triggered by
urgency itself, such as congestion, mobility, or sudden load surges. To minimize delay,
systems may perform partial state transfer or skip expensive verification steps\cite{p2cle}.

This weakens state provenance and exposes UC pipelines to rollback attacks, forked
state versions, or subtle state poisoning during handoff. The danger in UC is rapid
amplification: once an inconsistent or poisoned state is accepted, it propagates quickly
through dependent tasks operating under tight deadlines. This effect can be observed
through provenance gaps, state-hash mismatches, or increased rollback events during
urgent execution. A UC-appropriate mitigation is to attach lightweight provenance tags
and hash chains to migrating state, enabling fast verification without introducing
heavy auditing overhead.

\paragraph{Property 3: Isolation guarantees under emergency multi-tenancy.}
Edge and fog nodes commonly execute multiple workloads using containers or virtual
machines. Under normal conditions, isolation policies limit interference and prevent
data leakage between tenants. During urgent operation, however, bursty demand and
limited resources increase co-location pressure, and isolation constraints may be
relaxed to preserve latency.

This creates several UC-relevant risks, including cross-tenant data leakage,
side-channel exposure, and resource interference that degrades real-time performance.
More critically, isolation failures can enable attackers to pivot from one urgent
workflow into others, compounding system-wide impact. These effects can be analyzed
by measuring co-location rates for sensitive tasks, the frequency of privileged
exceptions, and observed interference during urgent episodes. From a UC perspective,
isolation must have a \emph{minimum acceptable floor}: latency optimization should
never collapse separation for high-impact workloads.

\paragraph{Property 4: Security-service starvation due to priority scheduling.}
To meet deadlines, UC platforms often rely on aggressive priority scheduling. While
effective for urgent tasks, this approach can starve security services such as monitoring,
attestation, anomaly detection, and logging, which are treated as background activities.
Under sustained urgent load, the system may enter periods where security mechanisms
are effectively inactive\cite{hasanp22017contegoadaptiveframeworkintegrating}.

This is particularly dangerous in UC, as attackers can exploit these windows to act
when detection capability is weakest, allowing compromises to spread before monitoring
recovers. This behavior can be analyzed by measuring how often security services are
deprived of CPU or network resources, along with increases in time-to-detect (TTD) and
time-to-contain (TTC). Rather than running heavyweight security continuously, UC
systems should reserve small but guaranteed ``security heartbeats'' that ensure essential
integrity checks remain active even during peak urgency.

\subsubsection{Identity Management at the Edge}
Identity management ensures that only authenticated and authorized entities can
participate in distributed computation. In Urgent Computing (UC), however, strict
deadlines and intermittent connectivity place direct pressure on identity workflows
at the edge. To reduce latency, authentication and authorization steps are often
simplified, cached, or temporarily relaxed. Because UC systems act on data and
decisions immediately, even brief identity failures can influence system behavior
before detection or recovery is possible.

We analyze this challenge through four identity-related security properties that
degrade most under urgency\cite{WANG2025111718}.

\paragraph{Property 1: Authentication strength under time pressure.}
Under normal operation, strong authentication mechanisms provide confidence in
entity legitimacy. In UC, these mechanisms are frequently reduced to low-latency
checks, such as cached or long-lived credentials. This increases the likelihood of
spoofing and replay attacks, particularly at the edge where monitoring is limited.
The impact is amplified in UC because injected data or commands are consumed
immediately by downstream components. This degradation can be analyzed through
authentication bypass rates and credential reuse frequency during urgent execution\cite{s25061649}.

\paragraph{Property 2: Authorization correctness and least-privilege erosion.}
Authorization systems are designed to enforce least privilege, but UC often introduces
emergency roles or broad permission overrides to avoid delays. These temporary
expansions weaken access boundaries and allow entities to perform high-impact actions
beyond their intended scope. In UC environments, such over-privileged access can
quickly propagate across the edge--cloud continuum. This effect can be assessed by
tracking emergency permission usage, scope breadth, and revocation delays\cite{10132479}.

\paragraph{Property 3: Identity binding to device and execution context.}
Reliable identity management depends on strong binding between an identity and its
underlying device or execution context. UC frequently involves rapid onboarding of
edge devices and ad hoc infrastructure, weakening this binding. As a result, cloned
identities or compromised nodes may inject untrusted data into urgent workflows.
This risk is particularly severe in UC systems where sensing, inference, and actuation
are tightly coupled. Identity binding failures can be analyzed through missing
attestation evidence and inconsistencies between identity claims and runtime posture\cite{Sadique2023DIdMEIoTD}.

\paragraph{Property 4: Auditability and non-repudiation under urgency.}
Identity management also relies on auditability and non-repudiation. During urgent
execution, logging and auditing are often reduced to conserve time and resources.
This limits accountability and delays forensic analysis, allowing identity-based
attacks to persist unnoticed. Indicators of this degradation include log drop rates
and missing identity-to-action traces during urgent episodes.

\subsubsection{Protocol Exposure and Communication Security} Urgent Computing (UC) systems rely on fast, continuous communication across the edge--fog--cloud continuum to support time-critical decision making. To minimize latency, communication protocols are often optimized for speed and availability, sometimes at the expense of strong security guarantees. In UC, these trade-offs are particularly dangerous because communication channels directly carry control signals, sensor data, and AI inputs that immediately influence system behavior. We examine four communication security properties that are most vulnerable under urgency. \paragraph{Property 1: Message freshness and replay protection.} Secure communication protocols typically enforce freshness through nonces, timestamps, or sequence numbers to prevent replay attacks. In UC environments, reduced handshake complexity, clock drift, and intermittent connectivity weaken these protections. As a result, previously valid messages may be replayed and accepted during urgent execution. The UC-specific impact is severe: replayed sensor readings or control messages can trigger repeated or incorrect actions before detection. This degradation can be analyzed by measuring nonce reuse, timestamp skew, and replay detection failures during urgent episodes\cite{Arfaoui2019PrivacyTLS13}. 

\paragraph{Property 2: Integrity of control-plane communication.} UC systems depend heavily on control-plane messages for orchestration, workload migration, and coordination across layers. To reduce latency, integrity checks on these messages may be simplified or selectively disabled. This exposes control channels to message tampering or injection. Because UC systems act immediately on control decisions, even small manipulations can redirect workloads, alter AI inputs, or disrupt coordinated response. Indicators include integrity verification failures and unexplained control-plane deviations during urgent execution\cite{q12article}. 

\paragraph{Property 3: Availability under congestion and denial-of-service.} Urgent events often generate traffic surges that stress communication channels. Security mechanisms such as handshakes, re-keying, or verification may be delayed or dropped under congestion. This increases vulnerability to denial-of-service attacks that exploit protocol overhead. In UC, communication delays translate directly into missed deadlines or incomplete situational awareness. This property can be analyzed by observing handshake latency, packet loss, and retransmission rates during peak urgency. 

\paragraph{Property 4: Authenticity of data consumed by AI pipelines.} AI-driven UC systems consume large volumes of data from distributed edge sources. When communication security is weakened, adversaries can inject forged or manipulated data that appears legitimate at the protocol level. Because AI inference and actuation are tightly coupled in UC, such data manipulation can rapidly influence global decisions. This risk can be assessed through missing data provenance, inconsistencies across redundant data sources, and delayed anomaly detection during urgent execution\cite{papernot2017practicalblackboxattacksmachine}.

\subsubsection{Secure Boot and Firmware Integrity} Secure boot and firmware integrity ensure that devices start from a trusted software baseline and that low-level code has not been tampered with. In Urgent Computing (UC), these guarantees are often weakened due to rapid provisioning, emergency deployment, and the need to scale edge and fog infrastructure quickly. When firmware verification is delayed or bypassed, compromises at the lowest system layers can silently undermine all higher-level security mechanisms. We highlight four firmware-related security properties that degrade under urgency. 

\paragraph{Property 1: Boot-time verification bypass under rapid deployment.} Secure boot mechanisms normally verify firmware and bootloaders before execution. During urgent deployment, devices may be brought online with verification disabled or deferred to reduce startup time. This creates an opportunity for persistent low-level malware to execute before any runtime security controls are active. In UC, such compromises are particularly dangerous because infected nodes immediately participate in time-critical workflows\cite{bootticle}. 

\paragraph{Property 2: Firmware update integrity and authenticity.} Firmware updates rely on cryptographic validation to ensure authenticity. Under urgent conditions, update processes may be simplified or performed over insecure channels to avoid delay. This increases the risk of installing malicious or corrupted firmware, which can remain undetected throughout urgent execution. The impact is amplified in UC because firmware-level compromise affects all subsequent tasks on the device\cite{firmminproceedings}. 

\paragraph{Property 3: Attestation coverage during urgent execution.} Runtime attestation helps verify that devices continue operating in a trusted state. In UC, attestation frequency is often reduced or skipped due to resource constraints and latency pressure. As a result, compromised devices may continue operating without detection. This degradation can be assessed by tracking attestation gaps and delayed verification during urgent workloads\cite{Abera2016CFLAT}. 

\paragraph{Property 4: Persistence and propagation of firmware-level attacks.} Firmware-level compromises are difficult to detect and remove. In UC systems, where devices are tightly coupled across the edge--cloud continuum, a compromised firmware image can propagate through shared images, updates, or provisioning pipelines. This allows low-level attacks to persist across urgent episodes and influence system-wide behavior.

\subsubsection{Secret and Credential Management}
Secret and credential management protects cryptographic keys, tokens, and sensitive
configuration data used across distributed systems. In Urgent Computing (UC),
emergency execution and rapid scaling place direct pressure on these mechanisms.
To avoid disruption, key rotation, revocation, and secret isolation are often delayed
or simplified, increasing the exposure window for compromised credentials.

We identify four credential-related security properties that degrade under urgency.

\paragraph{Property 1: Key freshness and rotation delay.}
Regular key rotation limits the impact of credential compromise. In UC, rotation is
frequently postponed to avoid interrupting time-critical workflows. As a result,
compromised keys remain valid longer than intended. This effect can be analyzed by
measuring key age and skipped rotation events during urgent execution\cite{keyasinproceedings}.

\paragraph{Property 2: Secret sprawl during rapid scaling.}
UC often requires fast provisioning of edge and fog nodes, leading to widespread
distribution of secrets across heterogeneous devices. This increases the attack
surface, as secrets may be exposed on weakly secured nodes. Secret sprawl can be
evaluated by tracking how many nodes hold a given credential and how often secrets
are replicated during urgent episodes.

\paragraph{Property 3: Secure storage limitations at the edge.}
Strong secret protection assumes the availability of secure storage such as hardware
security modules or trusted execution environments. Many edge devices lack such
capabilities, and UC deployments may bypass secure vaulting for speed. This raises the
risk of credential leakage through memory or filesystem access\cite{206170}.

\paragraph{Property 4: Revocation latency and stale credentials.}
Timely revocation is essential once a credential is compromised. In UC, network
congestion and distributed caching can delay revocation propagation, allowing stale
credentials to remain accepted across layers. This risk can be assessed by measuring
revocation delay and stale-token acceptance during urgent execution.

\subsubsection{Zero-Trust Enforcement} Zero-Trust security assumes that no component is inherently trusted and that access decisions are continuously verified based on identity, context, and system state. In Urgent Computing (UC), however, strict deadlines and rapid cross-layer coordination directly challenge continuous verification. To meet urgency requirements, Zero-Trust checks are often relaxed, cached, or temporarily bypassed, creating short but critical trust gaps. We highlight four Zero-Trust properties that degrade under urgent execution. 

\paragraph{Property 1: Continuous verification under deadline pressure.} Zero-Trust relies on frequent re-authentication and policy re-evaluation. In UC, continuous verification is often reduced to avoid added latency. As a result, entities remain trusted longer than intended, even as their execution context changes. This allows compromised components to continue operating during urgent workflows before revocation occurs\cite{NIST800207}. 

\paragraph{Property 2: Context-aware access decisions.} Zero-Trust policies depend on up-to-date context such as device posture, location, and runtime behavior. Under urgency, this context may be stale, incomplete, or ignored to prevent delays. In UC, outdated context can lead to incorrect access decisions that propagate rapidly across the edge--cloud continuum. 

\paragraph{Property 3: Policy enforcement consistency across layers.} Effective Zero-Trust requires consistent enforcement from edge to cloud. UC systems, however, often apply asymmetric policies, with weaker enforcement at the edge to preserve responsiveness. These inconsistencies create attack paths that enable lateral movement across layers during urgent execution\cite{REN20251593}. 

\paragraph{Property 4: Containment of lateral movement.} Zero-Trust aims to limit lateral movement by enforcing strict segmentation and verification at each step. In UC, temporary trust relaxations and fast orchestration can unintentionally enable rapid lateral movement before anomalies are detected. Because UC systems are tightly coupled, such movement can escalate quickly into system-wide compromise.

\subsubsection{Cryptographic Key Distribution}
Under urgent conditions, inconsistencies in cryptographic key distribution create
trust asymmetries across layers. These inconsistencies can lead to authentication
failures or unauthorized access, disrupting coordinated urgent responses.

\subsubsection{Monitoring and Runtime Attestation}
High-frequency urgent workloads reduce the fidelity of monitoring and runtime
attestation. As a result, anomaly detection is delayed, allowing attacks to escalate
before corrective actions can be applied.

\subsubsection{Failure Propagation Across Layers}
Security failures originating at the edge propagate more rapidly in UC systems due
to tight coupling and AI-driven orchestration. Localized compromises can therefore
escalate into system-wide urgency violations.

\subsubsection{Effect on Urgency Guarantees}
Each weakened security property directly increases the likelihood of deadline
misses, decision uncertainty, and unsafe autonomous actions, undermining the core
guarantees of Urgent Computing.

\noindent
In Urgent Computing, edge--cloud security failures are not isolated events but
time-amplified faults that propagate across layers and directly degrade the
correctness and timeliness of AI-driven decisions.

Overall, security mechanisms in Urgent Computing degrade in tightly coupled ways under time pressure rather than failing in isolation. Local weaknesses rapidly propagate across the edge–fog–cloud continuum, amplified by AI-driven orchestration. Consequently, security failures in UC directly violate urgency guarantees by undermining correctness, timeliness, and safe autonomous decision-making.

\subsection{Use-Case–Driven Requirements in Urgent Computing Systems}
Across the industrial, infrastructure, and AI-driven use cases discussed in this chapter, Urgent Computing (UC) systems are consistently defined by strict requirements on latency, reliability, and safety, which fundamentally distinguish them from conventional non-urgent computing systems\cite{7053}.

\subsubsection{Latency Requirement }

From the use cases, latency represents the maximum allowable time between data acquisition, AI-based decision making, and actuation before the decision becomes ineffective or harmful. 

In industrial safety monitoring, emergency response coordination, and real-time infrastructure control, latency is bounded in milliseconds to seconds, as delayed decisions directly translate to physical damage, financial loss, or safety hazards. 

Failure in UC systems occurs when latency exceeds this bound even once, resulting in missed interventions, incorrect emergency actions, or uncontrolled system states. 
In contrast, non-urgent systems tolerate delayed execution, where increased latency primarily degrades performance or user experience rather than causing irreversible consequences.

\subsubsection{Reliability Requirement }

From the use cases, reliability denotes the system’s ability to consistently produce correct and timely outputs under dynamic conditions, partial failures, and peak load situations. 

UC workloads demand high reliability because decisions are often non-repeatable and must remain correct under stress, degraded inputs, or incomplete system visibility. 
Failure in UC systems manifests as a single incorrect or unavailable decision during a critical window, which can cascade into system-wide instability or unsafe operations. 

In non-urgent systems, reliability failures typically allow retries, fallbacks, or delayed correction without immediate or severe impact.

\subsubsection{Safety Requirement}

From the use cases, safety refers to the assurance that AI-driven actions do not cause physical harm, regulatory violations, or uncontrolled behavior, even when operating autonomously under time pressure.

 In UC scenarios, safety is tightly coupled with correctness and timing, as automated decisions often directly influence physical systems or human operators.
 
Failure in UC systems occurs when unsafe actions are taken faster than humans can intervene, or when explainability and verification are bypassed to meet deadlines.

\FloatBarrier
\section{Summary}
This Chapter examined the dual nature of industrial smartness enabled by modern AI across the IIoT–edge–cloud continuum, with particular emphasis on systems operating under urgent computing (UC) constraints. While AI-driven automation has significantly improved efficiency, responsiveness, and decision-making in industrial environments, our analysis demonstrates that these benefits are accompanied by amplified risks when systems operate under strict time deadlines and limited opportunities for recovery. In UC settings, failures are no longer isolated software faults or performance degradations; they are time-critical events that can rapidly escalate into safety hazards, physical damage, or large-scale service disruption.

Our findings show that the tight coupling and real-time execution inherent to UC architectures fundamentally alter the security landscape. Vulnerabilities such as weak sensor trust, fragmented trust continuity across layers, insecure edge execution, and latency-optimized but fragile pipelines become particularly dangerous when decisions must be made within milliseconds. In such environments, traditional security assumptions—such as delayed human intervention, retries, or offline correction—no longer hold. Instead, even minor disruptions or misclassifications can propagate across the device–edge–cloud continuum and directly influence physical actuation.

Beyond cyber threats, the side effects of AI failures in UC systems—such as false positives, missed critical events, overconfident decisions, and cascading failures—underscore the need for urgency-aware design and evaluation. These effects are not confined to individual components but span sensing, computation, orchestration, and human interaction layers. As a result, the sustainability of smart industrial solutions depends not only on advanced AI models but also on defense-in-depth architectures, cross-layer trust management, and continuous monitoring designed explicitly for time-critical operation.

It is important to acknowledge that this study focuses on current technologies and known threat models; future AI paradigms and tighter automation loops may introduce new urgency-driven risks that are not yet fully understood. Nevertheless, the key implication remains clear: in urgent computing environments, security is inseparable from time. Addressing vulnerabilities, strengthening protocols, and enforcing robust security controls is not merely a matter of long-term resilience, but an immediate operational necessity. Ensuring that AI-driven industrial systems remain safe, trustworthy, and effective requires treating urgency as a first-class design and risk assessment dimension—because in UC systems, delayed decisions often mean irreversible outcomes.
\section{Future Directions}
Beyond the ongoing increase in automation, future industrial systems are expected to integrate emerging technologies such as large language models (LLMs), agentic AI, and autonomous multi-agent workflows. While these advancements promise greater adaptability and decision-making capabilities, they also expand the spectrum of potential threats—introducing risks that may be more complex and less predictable than those observed today. 

To safeguard against these side effects, future research should focus on designing AI models and industrial architectures that are inherently secure, resource-efficient, and capable of dynamic threat mitigation. This includes the development of self-healing and anomaly-aware systems, privacy-preserving AI techniques, and standardized governance frameworks that ensure responsible adoption. Furthermore, security strategies must extend across both information technology (IT) and operational technology (OT) layers, enabling end-to-end resilience in next-generation smart industries.



\bibliographystyle{IEEEtran}
\bibliography{sn-bibliography}

\end{document}